\documentclass[a4paper,fleqn,usenatbib]{mnras}

\usepackage{newtxtext,newtxmath}

\usepackage[T1]{fontenc}
\usepackage{ae,aecompl}

\usepackage{graphicx}	
\usepackage{amsmath}	
\usepackage{amssymb}	

\title[Dynamical evolution of three-planet systems]{Understanding the origin of white dwarf atmospheric pollution by dynamical simulations based on detected three-planet systems }

\author[R. F. Maldonado et al.]{
R. F. Maldonado,$^{1}$, E. Villaver $^{2,3}$, A. J. Mustill $^{4}$, M. Chavez $^{1}$, E. Bertone $^{1}$  \\
\thanks{E-mail: raulfms@inaoep.mx}
$^{1}$Instituto Nacional de Astrof\'isica, \'Optica y Electr\'onica, Luis Enrique Erro 1, Tonantzintla, 72849, Puebla, M\'exico\\
$^{2}$Departamento de F\'isica Te\'orica, Universidad Aut\'onoma de Madrid\\
$^{3}$Centro de Astrobiolog\'ia (CAB, CSIC-INTA), ESAC Campus Camino Bajo del Castillo, s/n, Villanueva de la Ca\~nada, 28692, Madrid, Spain\\
$^{4}$Lund Observatory, Box 43, SE-22100 Lund, Sweden\\
}
\date{Accepted XXX. Received YYY; in original form ZZZ}

\pubyear{2020}

\begin{document}
\label{firstpage}
\pagerange{\pageref{firstpage}--\pageref{lastpage}}
\maketitle


\begin{abstract}
Between 25 -- 50\,$\%$  of white dwarfs (WD) present atmospheric pollution by metals, mainly by rocky material, which  has been detected as gas/dust discs, or in the form of photometric transits in some WDs. Planets might be responsible  for scattering minor bodies that can reach stargazing orbits, where the tidal forces of the WD can disrupt them  and enhance the chances of debris to fall onto the WD surface. The planet--planet scattering process can be triggered by the stellar mass-loss during the post main-sequence  evolution of planetary systems. In this work, we continue the exploration of the dynamical instabilities that can lead to WD pollution. In a previous work we explored two-planet systems found around main-sequence (MS) stars and here we extend the study to three-planet system architectures.  We evolved 135  detected three-planet  systems  orbiting MS stars to the WD phase by scaling their orbital architectures in a way that their dynamical properties are preserved  by using the $N$-body integrator package \textsc{Mercury}. We find that 100 simulations (8.6\,$\%$) are dynamically active (having planet losses, orbit crossing and scattering) on the WD phase,  where low mass planets (1--100 $\mathrm{M}_\oplus$) tend to have instabilities in Gyr timescales while high mass planets ($>$ 100 $\mathrm{M}_\oplus$) decrease the dynamical events more rapidly as the WD ages.  Besides, 19 simulations (1.6\,$\%$) were found to have planets crossing the Roche radius of the WD, where 9 of them had planet--star collisions. Our three-planet simulations have an slight increase percentage of simulations that may contribute to the WD pollution than the previous study involving two-planet systems and have shown  that planet--planet scattering is responsible of sending planets close to the WD, where they may collide directly to the WD, become tidally disrupted or circularize their orbits, hence producing pollution on the WD atmosphere.

\end{abstract}




\section{Introduction}

Metallic lines (Mg, Si, Fe...) have been found in the ultraviolet--optical spectra of nearly 25\,$\%$ to 50\,$\%$ of all white dwarfs (WDs) \citep{zuckerman2003,koester2014,harrison2018,wilson2019}. Given that the gravitational settling time for these WDs is orders of magnitude shorter than their cooling time  ($t_\mathrm{cool}>100$ Myr for WDs with $T_\mathrm{eff}<$ 20000 K)  it is not expected to see elements heavier than hydrogen and helium in their spectra (see e.g.  \citealt{fontaine1979,paquette1986,wyatt2014}). Thus, cool metal-polluted WDs must be currently accreting material from their surroundings \citep{farihi2009,farihi2010,koester2014}. 

To explain atmospheric pollution on WDs, rocky bodies are dynamically delivered to the WD's proximity where tidal forces destroy them, producing all the observed phenomena. These include i) near-infrared excesses and double-peaked features in the optical spectra which imply the presence of a dust/gas disc located at few solar radii from the WD surface \citep[e.g.,][]{kilic2007,gansicke2006,melis2012,wilson2014,guo2015};  ii) spectral signatures of metallic elements that originate from rocky material with a composition similar to that of the bulk Earth \citep{gansicke2012,jura2014,xu2014,harrison2018,doyle2019} that could originate from planet--planet collisions \citep{melis2017,bonsor2020}; iii) asteroid material detected as variable transits or emission features around WD 1145+017, ZTF J013906.17+524536.89,SDSS J122859.93+104032.9 \citep{vanderburg2015,manser2019,vanderbosch2020}. Additionally, indirect evidence of a planet orbiting at distances of less than 0.1 au from a WD has been inferred from the gas disc detected around WD J091405.30+191412.25 whose composition resembles that of an ice giant planet's atmosphere \citep{gansicke2019,veras2020}.   

As a star becomes a white dwarf, it loses a considerable fraction of its mass. This means that the planet:star mass ratio increases, which can radically change the dynamics and stability of a planetary system. \cite{duncanlissauer98} demonstrated that the outer planets of the Solar System will remain stable following the Sun's mass loss, but \citet{debes2002} were the first  to explore the effects of mass loss on the dynamical evolution of generic planetary systems with two and three planets, using planets with equal masses and zero eccentricities. They studied how the planetary orbits evolve, finding that the boundaries of dynamical stability change as the star loses mass while becoming a WD. Then, minor bodies such as asteroids or even planets which first survive the main-sequence (MS) phase of the host star can become unstable at the WD phase. This seminal work has been extended by several studies, involving different number of planets and planet architectures in their simulations. One-planet \citep{bonsor2011,debes2012,frewen2014}  and two-planet systems \citep{smallwood2018} have been explored simulating the interaction of the planets with a planetesimal belt. They find that the delivery of sufficient quantities of material to the WD can only be achieved if the particle belt is very massive and the planet moves around the WD with a highly eccentric orbit.   \citet{voyatzis2013,veras2013, veras2013b,mustill2014,mustill2018} have expanded the dynamical studies of the evolution of two- \citep[adding tidal interaction with the star,][]{ronco2020}, three-, four-planet systems \citep{veras2016b}, and even ten-planet systems  \citep{veras2015} exploring a wider range of planet masses and orbital parameters. The overall conclusion is that instabilities that lead to the loss of a planet in two-planet systems can not explain the high incidence of atmospheric pollution observed in WDs, even when considering instabilities such as orbit crossing and orbital scattering. We reach a similar conclusion in \citet[][Paper\,I from now onwards]{maldonado2020}  where we expand the previously explored parameter space by using the planetary architectures of the two-planet systems found orbiting MS stars. Planets with different masses have in principle produced more interesting results. \citet{mustill2018} using three planets with three masses in the ranges between 1--30, 10--100 and 100--1000 $\mathrm{M}_\oplus$ found that low-mass planets are more efficient at delivering planetesimals toward the WD and for a longer time, in good agreement with the observed pollution trends.

Following the idea that a high multiplicity of planets can increase the planet--planet scattering events leading to the loss of a planet, we here study the dynamical evolution of three-planet systems using the same approach as in Paper\,I for two-planet systems.  In this paper, we expand the previously explored parameter space by using the planetary architectures of the detected systems orbiting MS stars. This parameter space is otherewise of high dimensionality. Using the observed planetary architectures of the three-planet systems found orbiting MS stars allows us to go beyond the parameter space explored by previous works---which typically used restricted masses, semimajor axis ranges, and low eccentricities---in a way informed by observations. In Section~\S2 we describe the simulations setup, in \S3 we present the results and in \S4 we discuss them, and finally in \S5 we summarize the conclusions of this work.

\section{Simulations}
\label{setup}

In order to build the architectures of the planetary systems we shall evolve, we have selected all the three-planet systems from the NASA Exoplanet Archive\footnote{See \citet{akeson2013}, https://exoplanetarchive.ipac.caltech.edu/ } and the Exoplanet Encyclopedia\footnote{See \citet{schneider2011}, http://exoplanet.eu/ } with reported discovery until June 2018 (3 systems, HD~125612, HD~136352 and HD~181433, were updated April 2020; see \S3).  We have excluded from the list one planetary system in which the host star is a pulsar, three binary systems of which two of them are catalogued as eclipsing binaries ({\it Kepler}-104, {\it Kepler}-114) according to the SIMBAD data base \citep{wenger2000}, and a third one (K2-136) which hosts a binary M type star companion on a $\sim40$\,au orbit \citep{ciardi2018} as the required treatment of these three-planet systems will be different from that of the rest of the simulations.

We obtained a final sample of 135 MS stars (we include 6 subgiants in this list) which each host three planets.  For those we select from the observations the planet and stellar masses, radius, and eccentricity, when available.  The mass range of the three-planet host star extends from 0.13 to 1.81 $\mathrm{M}_\odot$ and we have re-scaled the planetary systems to a 3 $\mathrm{M}_\odot$ mass star, keeping them dynamically analogous  as the original system. As explained in Paper\,I, the motivation behind the choice of evolving a 3 $\mathrm{M}_\odot$ is two-fold: i) polluted WDs have shown a mean mass of $\sim$ 0.7 $\mathrm{M}_\odot$ \citep{koester2014} corresponding to a progenitor star on the MS of 3 $\mathrm{M}_\odot$ and ii) the relatively rapid evolution of the host star model allows us to do a large number of simulations up to 10 Gyr in a reasonable computational time (a 3 $\mathrm{M}_\odot$ star lives 377 Myr on the MS and enters the WD phase after 477 Myr). The host star is evolved using the SSE code \citep{hurley2000} which considers the isotropic mass loss during the Red Giant Branch (RGB) and Asymptotic Giant Branch (AGB) phases. We set the Reimers $\eta$ mass loss parameter to be $\eta=0.5$  and adopted solar metallicity for the host star.

We use the  \textsc{Mercury} package \citep{chambers1999} modified by
\citep{veras2013b,mustill2018}, which takes into account the change of the stellar mass and radius along the different evolution phases. We implemented the RADAU integrator with a tolerance parameter of 10$^{-11}$ as in Paper\,I and in \citet{mustill2018}. Planets are removed from the simulations when they reach an orbital distance of $1\times10^6$ au from the central star which we consider an ejection. Planets colliding with each other or with the stellar radius are also removed.

\label{sec:scaling}

To conserve the Hill stability criterion (see Section~\ref{equa} below) in the simulations with the adopted  $3\mathrm{\,M}_\odot$ star, we multiply the mass of each planet by a scale factor defined as $f=3\mathrm{\,M}_\odot/M_*$, where $M_*$ is the observed mass of the host star in the system. 

\subsection{Planet Masses and Radii }

\begin{figure}
\begin{center}
\begin{tabular}{cc}
\includegraphics[width=8cm, height=6.5cm]{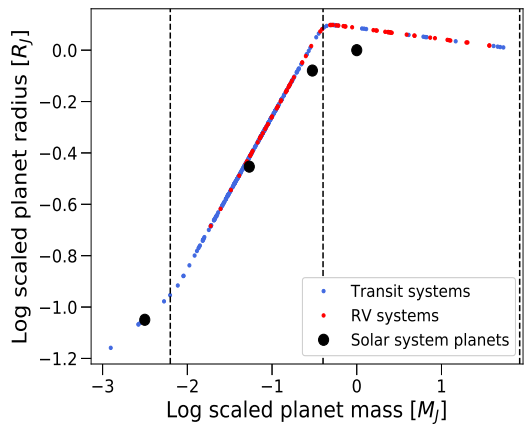} \\
\end{tabular}
\caption{Mass-radius relation of the scaled parameters of the planets in our three-planet systems using the FORECASTER package of \citet{chen2017}. Different colours represent the detection method (blue transit planets and red RV planets). From left to right the dashed vertical lines mark the different mass regime limits  of the Earth, Neptune and Jupiter mass planets respectively. The  Earth, Neptune, Saturn and Jupiter mass-radius location is shown as well as black dots.}
\label{mr}
\end{center} 
\end{figure}

To determine the physical parameters of the planetary system that are not available from the observations (i.e. mass of the planet when it is detected by transit and has no radial velocity (RV) measurement) we have used the PYTHON package FORECASTER \citep{chen2017}  and, as in Paper\,I, assume that the standard deviation in the input parameter for the mass and radius is 5\,$\%$ and use the median after 100 runs. Then, after scaling the masses of the planets, we proceed to recalculate the radius for each planet using FORECASTER. In Fig. \ref{mr} we show the mass-radius relation of our sample of planets, scaled by the factor $f$.   
In red we show the planets with RV measurements (RV systems) and in blue the transiting planets (Transit systems). The vertical dashed lines depict the mass limits proposed in \citet{chen2017} for Terrestrial, Neptunian, and Jovian planets respectively. Black dots mark the location in the relation of Earth, Neptune, Saturn and Jupiter, for reference. Note that for RV detections the planet mass is always assumed to be $m=m\sin i$.

\subsection{Initial Orbits}

\label{equa}

We placed the innermost planet (planet\,1) at a semimajor axis $a_0$ = 10\,au from the star so that the tidal forces are negligible during the RGB and AGB phases. We adopted this distance because \textsc{Mercury} does not include tidal forces and to allow a more direct comparison with previous works. This distance is justified also in \citet{villaver2009,mustill2012}, who show that planets beyond 10\,au do not experience any tidal orbital decay. As discussed in Paper\,I, since the Hill stability limit depends directly on the semimajor axis ratio between the planets, we must preserve the ratios of the observed planetary system in our simulations. Thus, planet\,2 is placed at a distance of $(a_2/a_1)a_0$ and the outermost planet (planet\,3) is placed at distance $(a_3/a_2)(a_2/a_1)a_0$, ensuring that the semimajor axis ratio among planet\,3, planet\,2 and planet\,1 is also conserved where $a_1$, $a_2$, and $a_3$ are the observed semimajor axes of planets\,1, 2 and 3 respectively.

\begin{figure*}
\begin{center}
\begin{tabular}{cc}
\includegraphics[width=8cm, height=6.5cm]{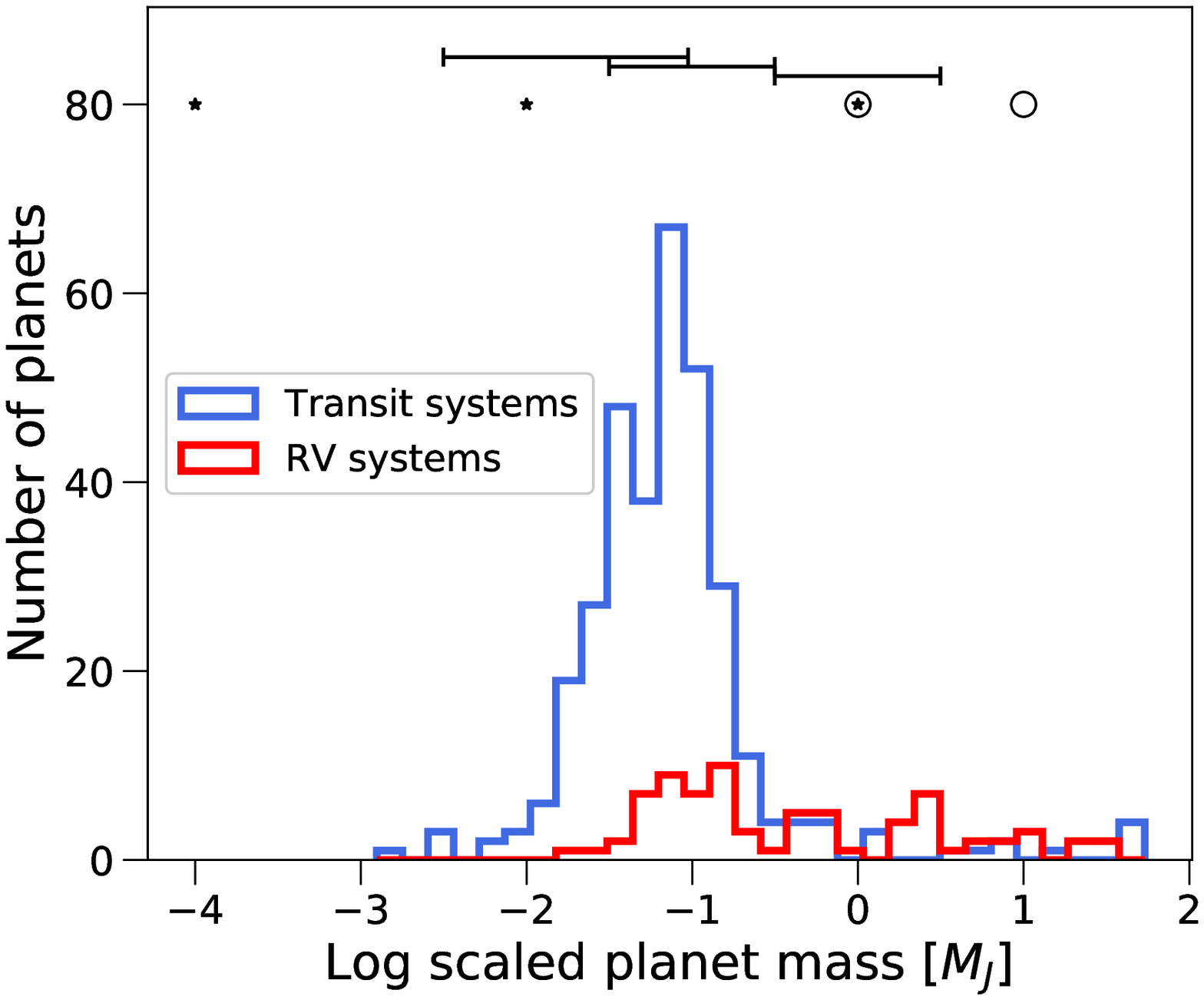}
\includegraphics[width=8cm, height=6.5cm]{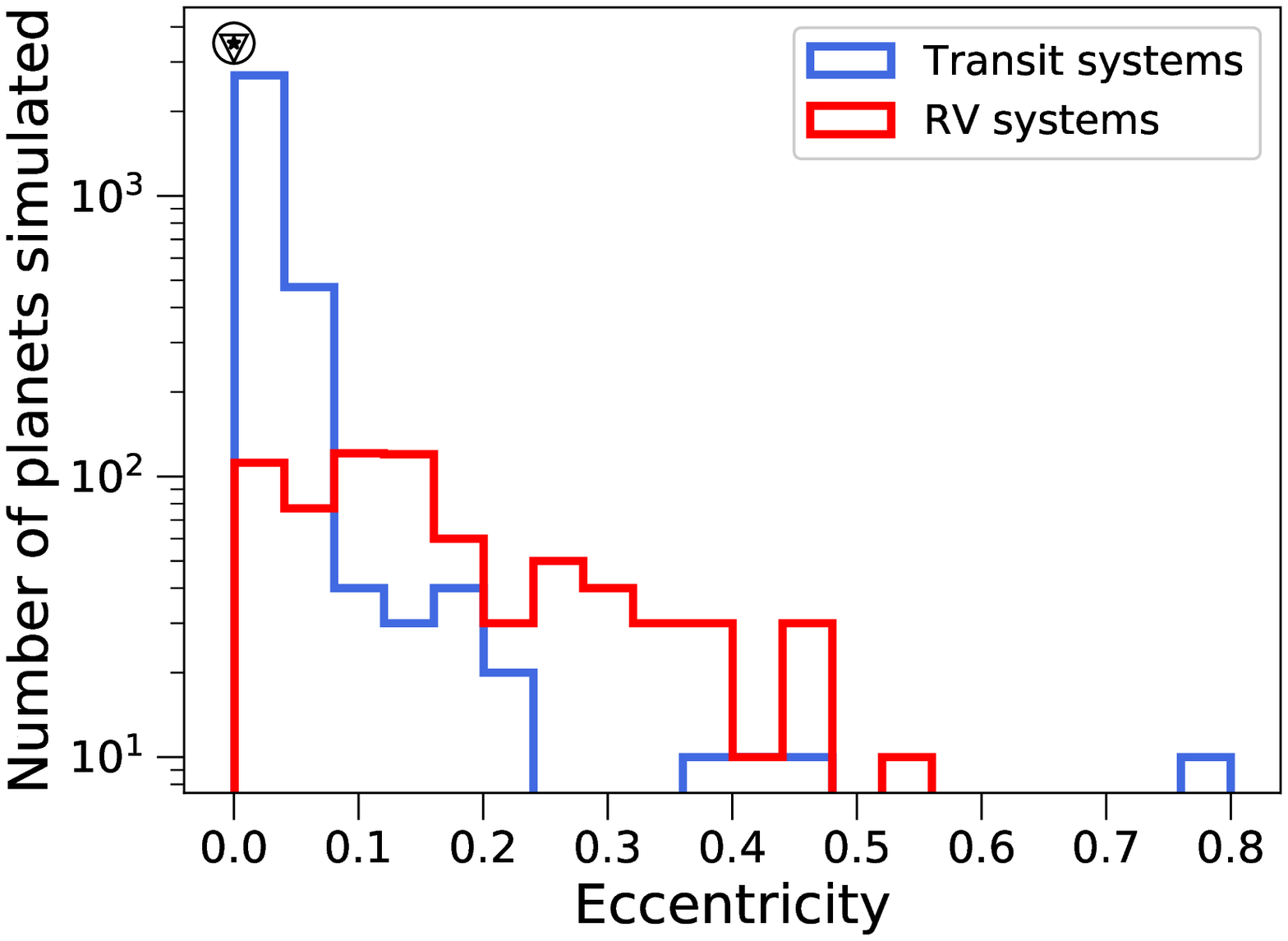} 
\end{tabular}
\caption{Scaled parameters of the planets considered in this work. Left, histogram of the scaled mass, showing 
in blue the Transit planets and in red the RV systems. Black starry symbols depict the masses used in \citet{debes2002} simulations, black circles the planet masses used in \citet{mustill2014} and in black horizontal lines the planet mass ranges explored in \citet{mustill2018} (1--30, 10--100, 100--1000 $\mathrm{M}_\oplus$).  The right panel shows  the distribution of eccentricity of all the simulated planets (4050). Colours and symbols are the same as in the left panel, with the addition of the inverted triangle for \citet{mustill2018} eccentricities in the upper part of the histogram. }
\label{4fig}
\end{center} 
\end{figure*}

The planet eccentricity and orbital inclination are also needed as input for \textsc{Mercury}. The eccentricity is taken from a Rayleigh distribution with a $\sigma$ parameter $\sigma=0.02$ \citep{pu2015} when it is not available in the catalogs: see \citet{vaneylen2015} and \cite{moorhead2011} for a justification of our choice. The inclinations are also randomly selected from a Rayleigh distribution with $\sigma = 1.12^\circ$  \citep{xie2016}. Choosing small but non-zero inclination angles is adequate for global stability studies \citep{veras2018}.

In the left panel of Fig. \ref{4fig} we show the histogram of the  scaled mass distribution of the planets we used in this work. We note that in this paper we cover a parameter space in planet mass that has not been covered in previous simulations (see the symbols at the top of the plot where we have marked the parameters used in previous works). The right panel of Fig. \ref{4fig} shows the distributions of eccentricity of the planets simulated in this work. Note that previous simulations done of three-planet systems have only used $e=0$ \citep{debes2002,mustill2014,mustill2018} so we are exploring a much wider parameter space in eccentricity here. We have not simulated systems at zero eccentricity because this often reflects a lack of information, and it is more realistic to simulate them using a small eccentricity with a Rayleigh distribution with $\sigma =0.02$.

The Rayleigh distribution assumed  for those planets that did not have eccentricity measurements in the catalogues is evident in the blue histogram. Higher eccentricities dominate the sample of planets with RV measurements.

In order to show the architecture of the three-planet systems in our simulations, we calculate the planet separation $\Delta$ in terms of mutual Hill radii for the planet pairs 1--2, 2--3 and 1--3 where the mutual Hill radius $R_\mathrm{m,Hill}$ is defined as
\begin{equation}
R_\mathrm{m,Hill}=\frac{a_i+a_j}{2}\left(\frac{m_i+m_j}{3M_*} \right)^{1/3}
\end{equation}
and the indices $i,j$ refer to the planet $i,j$ for $i,j=1,2,3$, the respective planets, $a$ is the semimajor axis, $m$ is the mass of the planets and $M_*$ corresponds to the host star mass on the MS, 3 $\mathrm{M}_\odot$. Note that  unlike the case of two-planet systems with the \emph{Hill stability limit} (see also \citealt{gladman1993,donnison2011,veras2013b}) for three-planet systems there is no analytic formulation from which we can determine if the system will be dynamically stable or the planets may have close encounters. Two main dynamical effects can result in the loss of a planet, the Hill and Lagrange instabilities. In Hill-unstable systems, planets are close enough to collide with each other or to cross each other's orbits. In Lagrange-unstable systems,  at least one planet is lost of the system via collision with the star or ejection outwards from the system. The Lagrange stability limit does not have any analytic formulation; thus, it can only be found by performing numerical simulations.

\begin{figure*}
\begin{center}
\includegraphics[width=18cm, height=11cm]{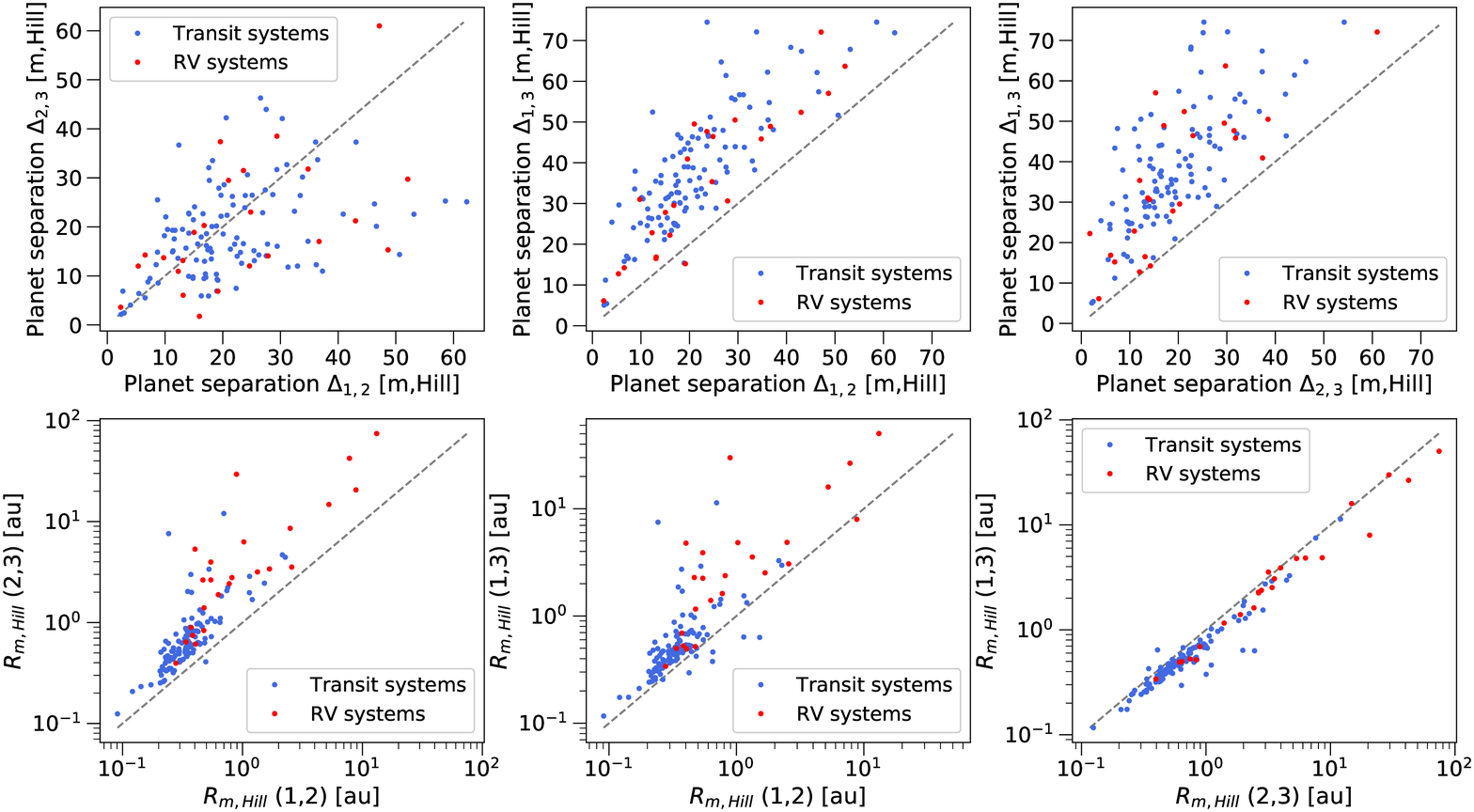} 
\caption{Top figures, planet separation $\Delta$ in mutual Hill units as a function of the planet separation of the different pairs of planets. In the bottom figures we display the mutual Hill radius for the three pairs of planets calculated with equation 1 using the 3$\mathrm{M}_\odot$ MS mass. The gray dashed line show the identity function and the colours are for the Transit (blue) and RV (red) systems.}
\label{deldel}
\end{center} 
\end{figure*}

In Fig. \ref{deldel} the three top panels show the planet separation  $\Delta =(a_j-a_i)/R_\mathrm{m,Hill}$ of different planet pairs $i,j$ as a function of the other planet pair. In the lower panels we show the same but for  $R_\mathrm{m,Hill}$ of each pair. Blue and red colours, as before, correspond to Transit and RV systems, respectively and the gray dashed line marks the location of the one-to-one relation. Regarding the planet separation from the top left panel we notice a large scattering in the separation between planets\,2 and 3 with respect to the distance between planets\,1 and 2. The middle and right top panels clearly indicate that the planet separation between the 1--3 pairs is usually larger than the distance between the two pairs of adjacent planets (the exceptions, in the top middle panel of Fig.~\ref{deldel}, have a very massive outer planet).  No distinction between Transit and RV systems is found in the top panels. On the other hand, the lower panels show that the planet pairs 2--3 and 1--3 have larger mutual Hill radius than the pair 1--2, which means that either planet\,2 or planet\,3 has a larger mass than planet\,1. The lower right panel shows that the Hill radius of the pair 1--3 is shorter that the pairs 2--3, since planets\,1 and 3 are more separated than planets\,2 and 3. In the three panels the dots follow a positive trend and mutual Hill radius for Transit planets is in general smaller than the mutual Hill radius of RV systems. 

The systems in Fig. \ref{deldel} span a broad range of separations in mutual Hill radii. Those spaced closely enough are expected to be destabilised by mass loss even when on circular orbits, as shown by \citet{mustill2014}. Here, we include wider systems than \citet{mustill2014}, which might be expected to remain stable; however, note that we include non-zero eccentricities for the planetary orbits that render systems significantly more unstable. Hence, we numerically integrate them all.

\begin{figure*}
\begin{center}
\includegraphics[width=16cm, height=14.5cm]{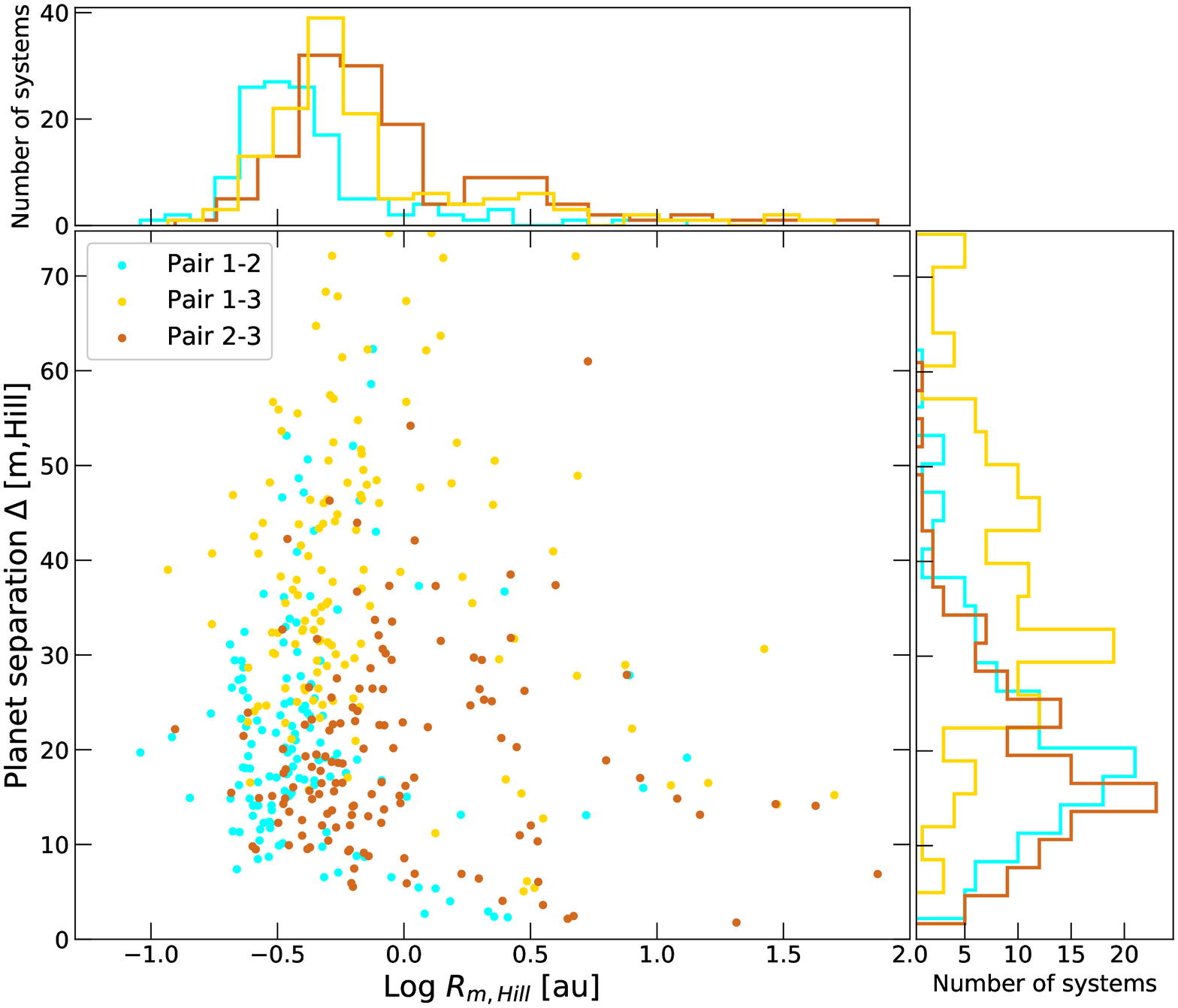} 
\caption{Scatter plot of the planet separation $\Delta$ as a function of mutual Hill radius of planet pairs 1--2, 1--3, 2--3 of our three-planet systems using the MS mass of the star. Light blue color depicts the planet pair 1--2, the yellow color shows the pair 1--3 and the brown represent pair 2--3. In the top panel histogram we display the mutual Hill radius distribution of the three-planet pairs as well as the planet separation $\Delta$ of each pair in terms of mutual Hill radius.}
\label{delhil}
\end{center} 
\end{figure*}

In Fig. \ref{delhil} we display a scatter plot of the planet separation $\Delta$ as a function of the mutual Hill radius of the three pairs of planets mentioned before. The scatter plot is accompanied by the respective histograms of the parameters in the top and left part of the panel. In light blue dots we show the pair 1--2, in yellow the pair 1--3 and in brown the pair 2--3. The top panel histogram show that planet\,1 and 2 have a mutual Hill radius smaller than the planet pairs 1--3 and 2--3, and the latter pairs display a similar distribution of Hill radius. Furthermore, the histogram of the planet separation $\Delta$ confirms that the $\Delta$ of pairs 2--3 is smaller than the other planet pairs, peaking around 15 mutual Hill radii while the separation between planet\ 1 and\ 2 peaks around 19  and the pair 1--3 peaks around 31 mutual Hill radii respectively.

\begin{figure*}
\begin{center}
\begin{tabular}{cc}
\includegraphics[width=18cm, height=6cm]{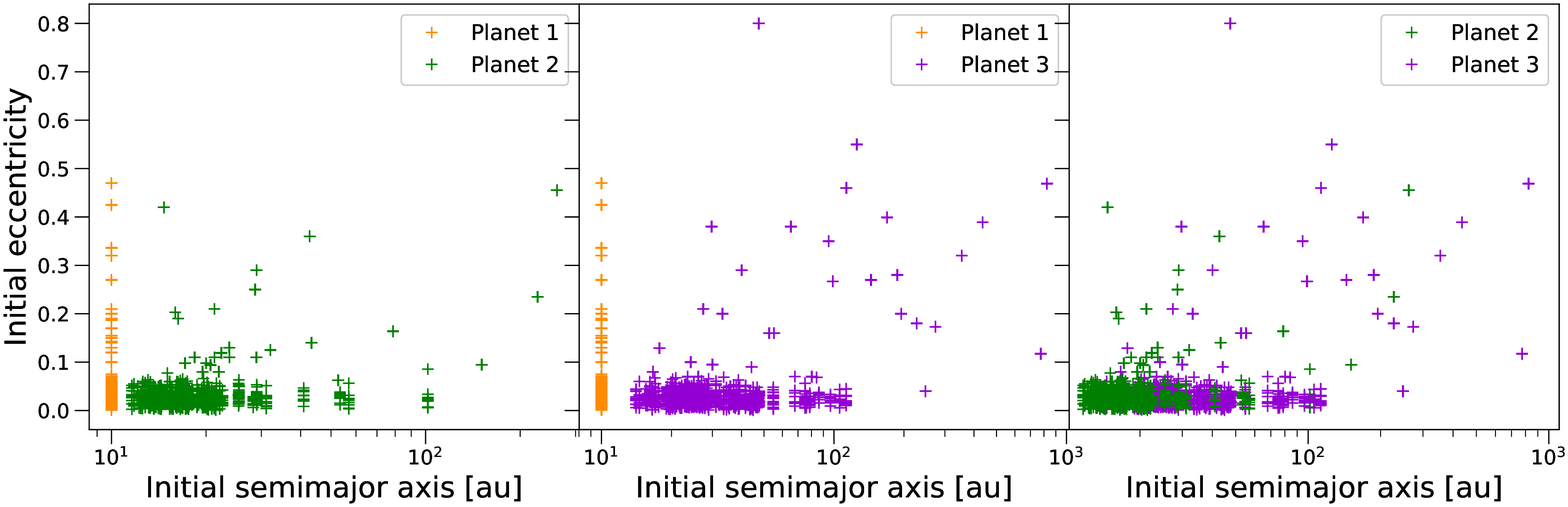} \\
\end{tabular}
\caption{Initial eccentricities as a function of initial semimajor axis of the three-planet systems simulated in this work. Orange, green and purple colours refer to planet\,1, 2 and 3 respectively. The three planets cover a wide range in eccentricities. Our planetary system configuration locates planet\,1 always at initial semimajor axis of 10 au while planet\,2 and 3 will have the semimajor axis which correspond to the semimajor axis ratio of each observed system.}
\label{initea}
\end{center} 
\end{figure*}

To finish our description of the parameter set-up, in Fig. \ref{initea} we show the initial semimajor axis and eccentricity of the scaled three-planet systems simulated in this work. We split it in panels to have a better look at the distribution of parameters in planet\,2 and planet\,3, since they overlap if they are displayed together. Orange, green and purple plus symbols refer to planet\,1, planet\,2 and planet\,3  respectively. All planets\,1 are located at 10 au, covering the eccentricity range from 0 to 0.5, planet\,2 and 3 are scattered and some of them are located at high eccentricities and large semimajor axis.

Finally, we proceeded to run 10 simulations per system configuration changing the inclination and the eccentricity, when unavailable, of the planet orbits using the Rayleigh distribution mentioned above. Additionally, the three orbital angles: the argument of the perihelion, the mean anomaly and the longitude of the ascending node of each planet are randomly selected from an uniform distribution of angles between 0 and 360$^\circ$ in each simulation. 

\section{Results}

\label{resu}
We have dynamically followed during 10 Gyr the evolution of 1350 planet configurations corresponding to 135 planetary systems (we run 10 simulations per system) of three planets orbiting a 3 $\mathrm{M}_\odot$ host star for which the MS lasts up to $t=$ 377.65 Myr; the RGB and AGB phases ranges between 377.65--477.57 Myr, and the WD phase starts at $t>$477.57 Myr. 

The planetary systems we are using as templates for our simulations are mostly discovered orbiting stars that are Gyr old. This means that the system has remained dynamically stable for a long period of time. Finding any early instability in our simulations, mainly in the MS phase, implies that something is wrong with the physical parameters reported in the catalogs. We find 25 three-planet systems that become unstable on the MS, either losing a planet or having orbit crossing events between the planets. 22 of these systems have at least 1 pair of planets close to mean motion commensurabilities of first and second order, mainly having period ratios within 10\,$\%$ of the 2:1 or 3:2  commensurability.  The planet pairs in three systems, namely HD~125612, HD~136352 and HD~181433 could not be associated to any mean motion commensurability, but, we found that \citet{ment2018,horner2019,udry2019} have recently updated the orbital solutions of these systems. We re-ran these systems with the updated and more stable solution and this time we found them to be stable on the MS.

As mentioned before, the innermost planet in our simulations is set to a distance of 10 au from the star to avoid the tidal forces during the giant phases and the rest are located at distances scaled so as to conserve the semimajor axis ratio. The planetary systems that are dynamically active and have orbit crossings would be more prone to have planet--planet collisions if they were in more compact configurations, enhancing the chances of losing a planet. Out of the 220 simulations with mean motion commensurabilities (the 22 systems mentioned above) we find that 122 of them lose planets on the MS phase.  Additionally, we find that orbit crossing is present on the MS phase in 2 simulations with a planet loss on the pre-WD phase (377--477 Myr), in 45 simulations with planet losses on the WD phase and in 21 simulations without any planet loss during the 10 Gyr of simulated time. In summary out of the 220 simulations with mean motion commensurabilities we find that 190 simulations have dynamical instabilities on the MS phase.  The other 30 just happen to have orbital configurations that keep them stable on the MS even with mean motion commensurabilities, without losing any planet or having orbit crossing on the MS. We have decided to exclude the 190 simulations that show some sort of instability on the MS (some of them eventually propagate into the WD phase) from the analysis that follows. A proper analysis of the stability of these systems requires a far more detailed study (e.g. in terms of the orbital angles of each planet initial set-up) than the one used here.  

We use 1160 (out of the 1350 computed) simulations to conduct our statistical analysis.  Thus, the numbers of planet losses on the different evolutionary stages are as follow: i) none on MS phase (we remove those simulations), ii) 2 simulations with a planet--star collision instability at the AGB tip and iii) 76/1160 (6.6\,$\%$) simulations where a planet is lost on the WD phase. The analysis of the 76 simulations where a planet is lost at the WD phase reveals that most of the planets are lost by ejection (71 of them) and only 9 of them are lost by a collision with the star (note that we have 2 simulations where two planets are lost by ejection on the same simulation and 2 simulations with the first planet ejected and the second one with a planet--star collision, leaving only one planet at the end of the simulated time). We find that no planet--planet collisions happen on the WD phase in any of our simulations. In Table \ref{tabwd} we present the statistics of the planet losses on the WD phase. The first percentage is with respect of the 1160 simulations and the second one is with respect to the number of planets simulated (3480).

\begin{table} 
\small
\begin{center}
\caption{Number of planet instabilities (collision between the planets, planet collision with the star, ejection) appearing at the WD phase. The first percentage is the fraction of simulations in which the given outcome occurred; the second, the fraction of planets experiencing it with respect to the total number of planets. Since 2 simulations lose the first planet by ejection and the second one by planet--star collision, we count separately the same simulation in ejection and planet--star collision.}
\label{tabwd}
\resizebox{.47\textwidth}{!}{%
\begin{tabular}{l r r }
\noalign{\smallskip} \hline \noalign{\smallskip}
 & \multicolumn{2}{c}{Planet instabilities during WD phase} \\
 & Systems & Planets \\
\noalign{\smallskip} \hline \noalign{\smallskip}
{\bf Ejections} & 69 (5.9\,$\%$)& 71 (2.0\,$\%$)  \\
{\bf Planet--star collisions} & 9 (0.8\,$\%$) & 9 (0.3\,$\%$)   \\
{\bf Planet--planet collisions} & --    & -- \\
{\bf Total} & 76 (6.6\,$\%$)& 80 (2.3\,$\%$)   \\
\hline
\end{tabular}}
\end{center}
\end{table}

\subsection{Orbital configuration and unstable systems}

In the following we analyze the orbital configurations that produce unstable systems in terms of the planet separation measured with different variables (semimajor axis ratios, $\Delta$ in mutual Hill units, planet mass and planet:star mass ratios, and eccentricity ratios of the different pairs). 

\begin{figure*}
\begin{center}
\includegraphics[width=18.5cm, height=6cm]{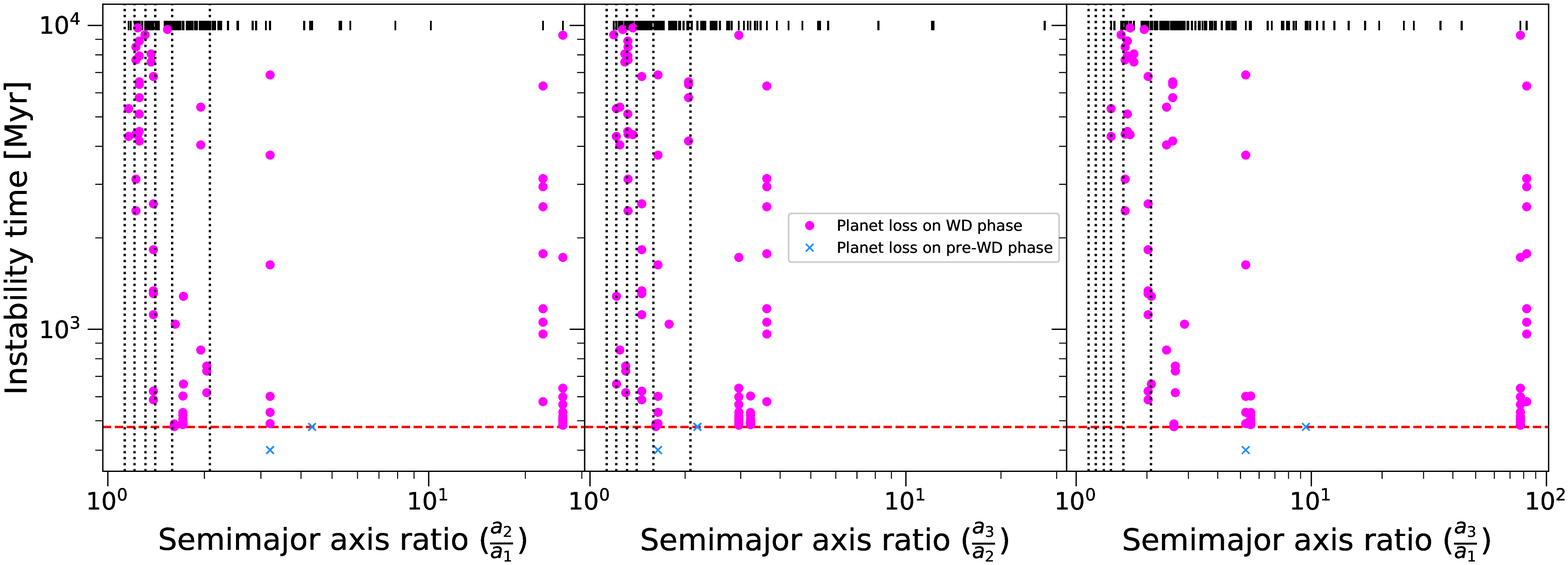} 
\caption{Instability time as a function of the semimajor axis ratio for different planet pairs. The black vertical ticks mark the $\frac{a_j}{a_i}$ semimajor axis ratio values of the three-planet systems used in this study. In blue x--shape symbols we show the planet losses before the WD phase while in pink dots we display the dynamical instabilities when the planets are lost on the WD phase. The red horizontal dashed line marks the time when the star becomes a WD. The location of the first and second order mean motion commensurabilities are shown as black dotted lines and are from left to right: 6:5, 4:3, 3:2, 5:3, 2:1, 3:1.}
\label{inar}
\end{center} 
\end{figure*}

In Fig. \ref{inar} we show the instability times as a function of the semimajor axis ratio $\frac{a_j}{a_i}$ of our three-planet systems. The times where a planet is lost on the WD phase are marked as pink dots, the blue crosses depict those systems that lose a planet before the WD (excluding the MS ones) and the black vertical lines shows all  the 1160 simulations considered in our statistics, most of them stable during the 10 Gyr simulated. In the left panel, we see that most of the instabilities where a planet is lost happen for semimajor axis ratios  $\frac{a_2}{a_1}$ $\leq$ 3.2. Two systems---HD~125612 and HD~181433---have $\frac{a_2}{a_1}>20$; in these cases, the innermost planet is (in the unscaled system) on an orbit of a few days' period, and the two outer planets at several au, separated from the inner one but comparatively close to each other. Indeed, when we look at the orbital proximity of the outer planet pairs in all of the systems that lose planets, we see that all the planet losses happen at semimajor axis ratio values  $\frac{a_3}{a_2}$ $\leq$ 3.6 with the largest one corresponding to the system HD~181433.

\begin{figure}
\begin{center}
\includegraphics[width=8.cm, height=6.5cm]{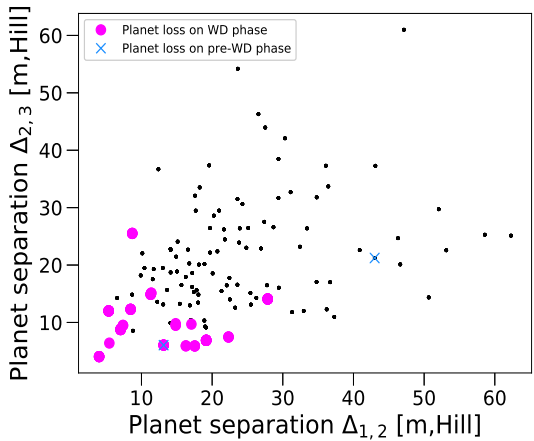} 
\caption{Planet separation $\Delta$ in mutual Hill radius units (computed using the MS  mass of the star) for different planet pairs. In black dots we mark the three--planet systems that do not loss any planet. In pink we show systems where their planets are lost on the WD phase, while the blue x-symbols are for the planets lost before the WD phase. }
\label{undel}
\end{center} 
\end{figure}

In order to understand the instabilities in terms of the different simulated parameters we present in Fig. \ref{undel} the $\Delta$ in mutual Hill units of the planet pair 2--3 as a function of the planet pair 1--2, where the black dots show the systems that do not lose any planet, pink dots are for those losing a planet in the WD phase, and the blue crosses are the systems losing planets in the pre-WD phase. Systems become unstable out to a maximum separation of $\Delta_{1,2}$ or $\Delta_{2,3}\approx28$, although the boundary is fuzzy, as in previous studies \citep[e.g.,][]{mustill2018}.

\begin{figure*}
\begin{center}
\includegraphics[width=18.5cm, height=6cm]{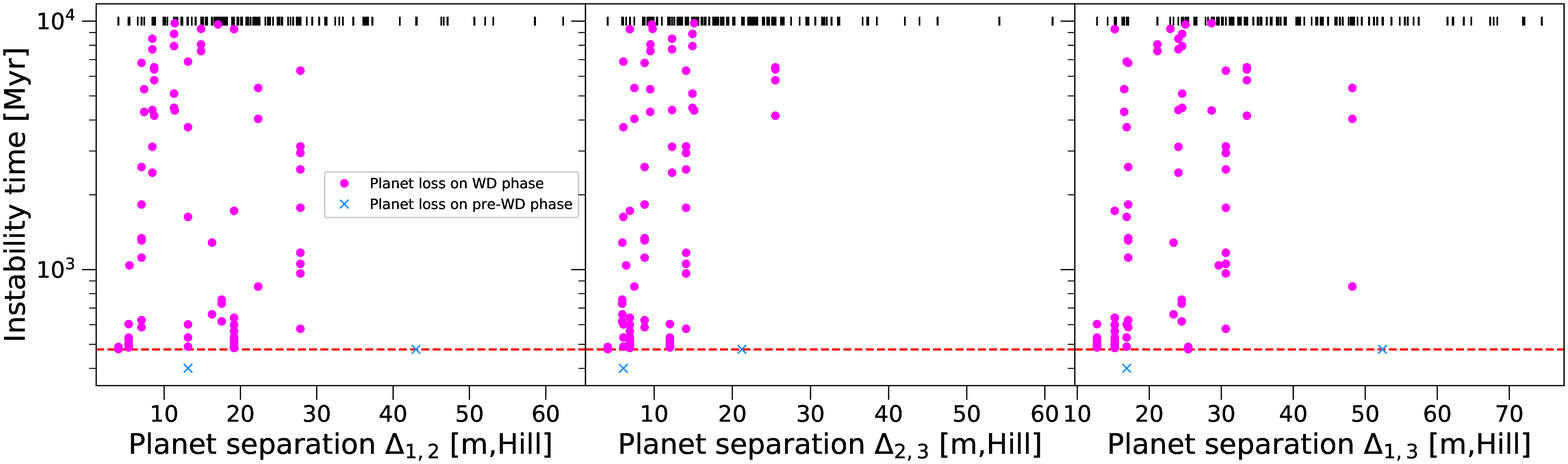} 
\caption{Instability time as a function of the planet separation $\Delta$ in mutual Hill radius units for different planet pairs calculated with the MS mass of the star. Symbols are as in Figure \ref{inar}.} 
\label{indel}
\end{center} 
\end{figure*}

These planet losses happen all along the WD phase as can be seen in Fig. \ref{indel} where we show the time at which planets are lost versus $\Delta$ of the planet pairs. The middle panel shows the more compact configuration for instabilities, that of planet pairs 2 and 3, where most of the instabilities are happening in $\Delta_{2,3}$ $\leq$ 15. Only one system ({\it Kepler}--26 with $\Delta_{2,3}=$ 25.5) has planet losses above this limit.

\begin{figure*}
\begin{center}
\includegraphics[width=18.5cm, height=7cm]{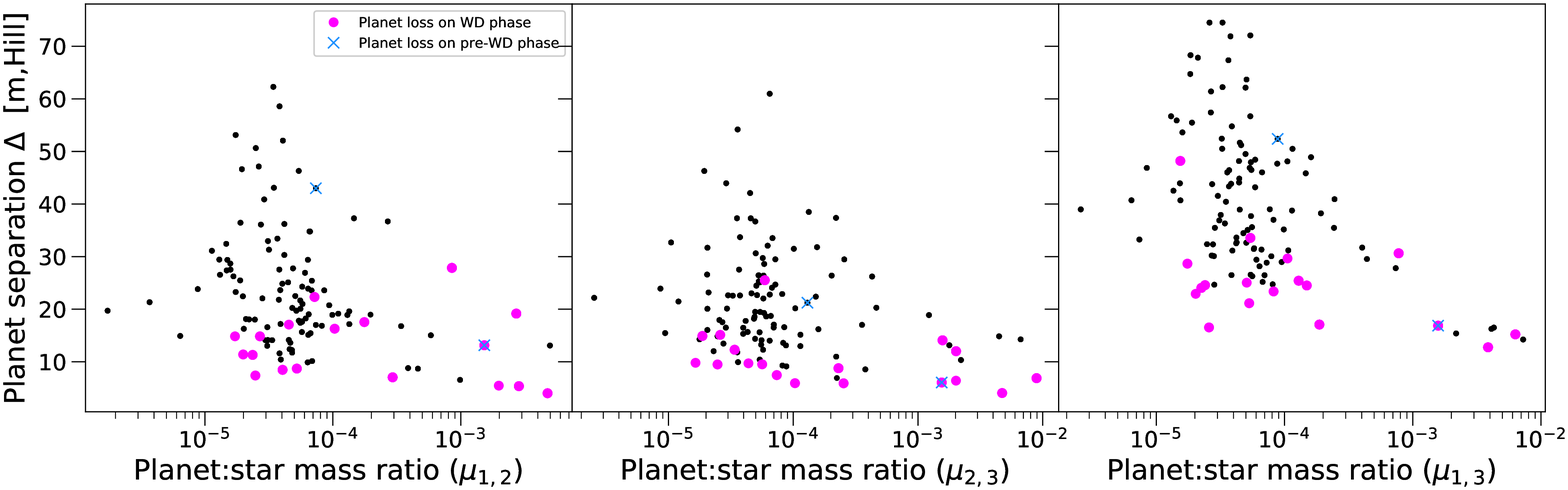} 
\caption{Planet separation $\Delta$ in mutual Hill radius units as a function of the planet:star mass ratio $\mu$ for the different planet pairs using the MS mass of the star. Symbols are as in Figure \ref{undel}.}
\label{mudel2}
\end{center} 
\end{figure*}

The distribution of the three-planet systems in the $\Delta$--$\mu$ space, where $\mu$ is the planet:star mass ratio defined as $\mu=\frac{m_i+m_j}{M_*}$ ($m_i$ and $m_j$ as the planet masses in the pair i--j and $M_*$ is the mass of the host star on the MS) is shown in Fig. \ref{mudel2} for the planet pairs 1--2, 2--3 and 1--3 from left to right, respectively. That figure shows that as the planet:star mass ratio increases, the required planet separation for an instability to take place is smaller.  The envelope where no dots are present is due to the relation between $\Delta$ and $\mu$ defined as, $\Delta_{\mathrm{max}}=2(\frac{m_1+m_2}{3M_*})^{-1/3}$ \citep[see][]{mustill2014}. We note that planet losses on the WD phase happen for very similar ranges for the pairs 1--2 and 2--3 (4.0 $\leq$ $\Delta_{1,2}$ $<$ 27.9 and 4.0 $\leq$ $\Delta_{2,3}$ $<$ 25.5), while for the pair 1--3 is slightly larger (12.7 $\leq$ $\Delta_{1,3}$ $<$ 48.2).

\begin{figure*}
\begin{center}
\includegraphics[width=18.5cm, height=6cm]{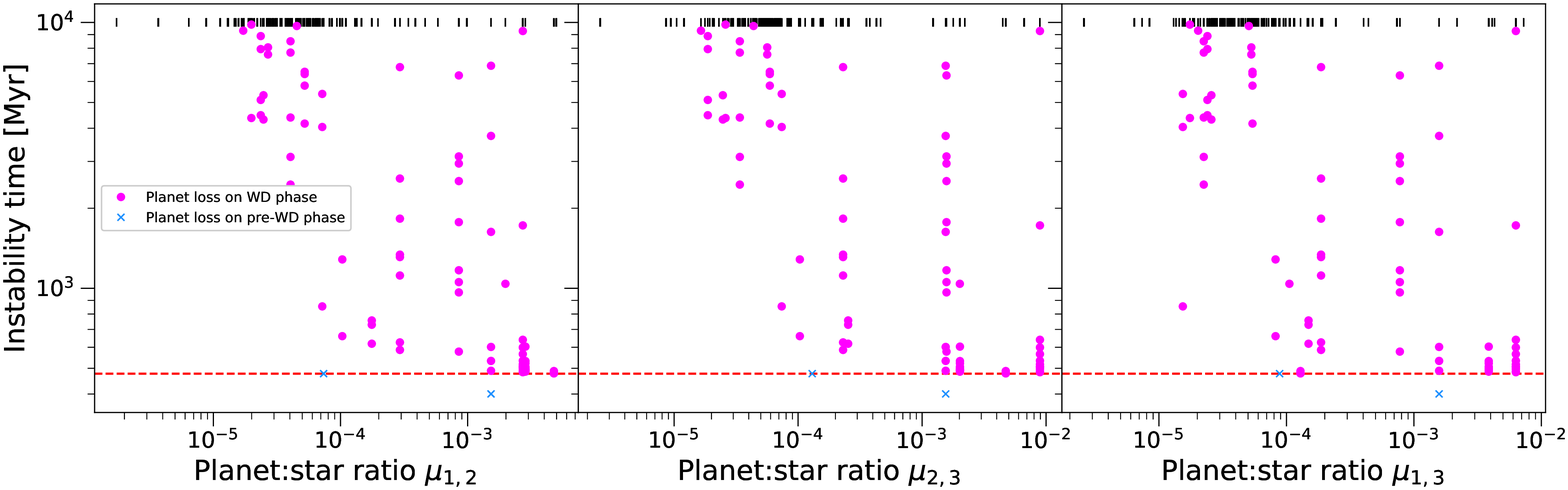} 
\caption{Instability time as a function of the planet:star mass ratio $\mu$ for different planet pairs using the MS mass of the host star. Symbols are as in Figure \ref{inar}.} 
\label{inmu}
\end{center} 
\end{figure*}

There is an slight trend that indicates that as $\mu$ decreases, the planet is lost at later times during the WD evolution, an effect more visible for the planet pairs 1--2 than 2--3 as can be seen in Fig. \ref{inmu} where as in Fig. \ref{indel} we show the instability times but this time as a function of the planet:star mass ratio $\mu$ for the different planet pairs.

\begin{figure*}
\begin{center}
\includegraphics[width=18.5cm, height=6cm]{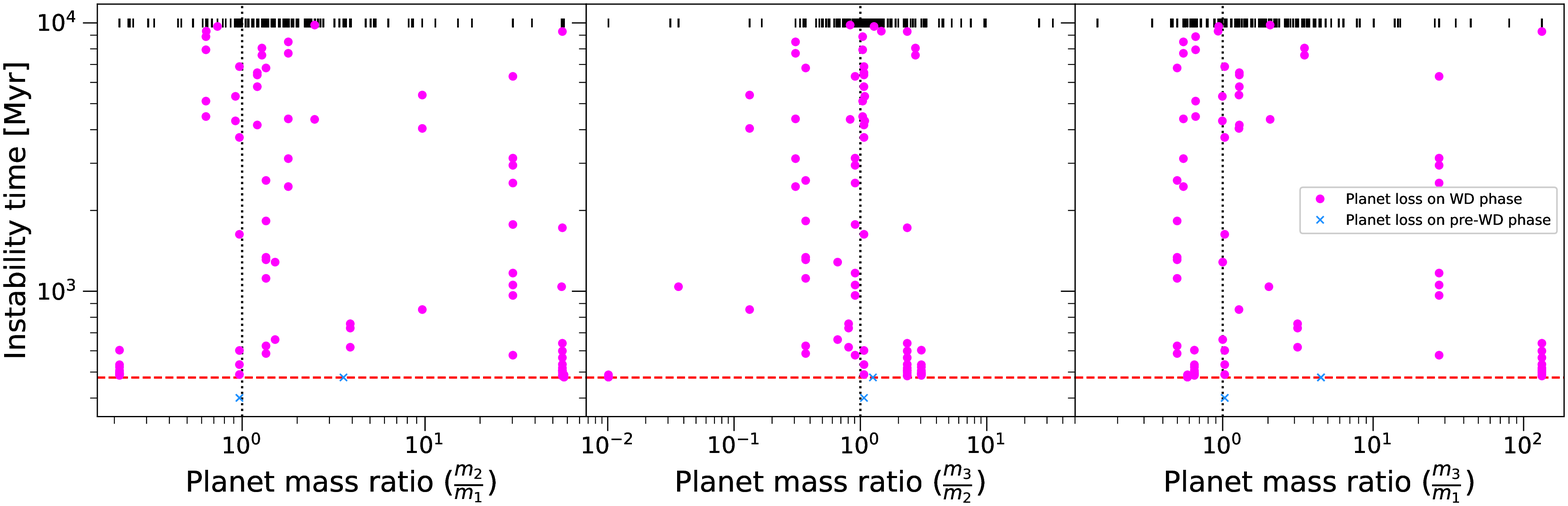} 
\caption{Instability time as a function of the planet mass  ratio for different planet pairs. The black vertical dotted line shows the equal mass planet ratio. Symbols are as in Figure \ref{inar}.} 
\label{inmr}
\end{center} 
\end{figure*}

In Fig. \ref{inmr} we show the instability times of planet losses with respect to the planet mass ratio of the different planet pair. From left to right are 1--2, 2--3 and 1--3, respectively. More than twice of the instabilities in the left panel have planet\,2 being more massive than planet\,1; planetary losses happen in 24 simulations with mass ratio $\frac{m_2}{m_1}$ $\leq$ 1 and in 52 simulations with $\frac{m_2}{m_1}$ $>$ 1, but note that $\frac{m_2}{m_1}$ $>$ 1 in 75\,$\%$ of the whole sample. On the other hand, we obtain quite similar number of simulations that are losing a planet if we compare the mass ratio of the planet pairs 2--3 and 1--3. We have 36 simulations with $\frac{m_3}{m_2}$ $\leq$ 1, 40 simulations with $\frac{m_3}{m_2}$ $>$ 1; 37 simulations  with $\frac{m_3}{m_1}$ $\leq$ 1 and 39 with $\frac{m_3}{m_1}$ $>$ 1. Planet\,3 is more massive than planet\,1 in 75\,$\%$ of the systems while planet pairs 2--3 are more equally distributed in mass ratio (56\,$\%$ of them have $\frac{m_3}{m_2}$ $>$ 1).

\begin{figure*}
\begin{center}
\includegraphics[width=18.5cm, height=6cm]{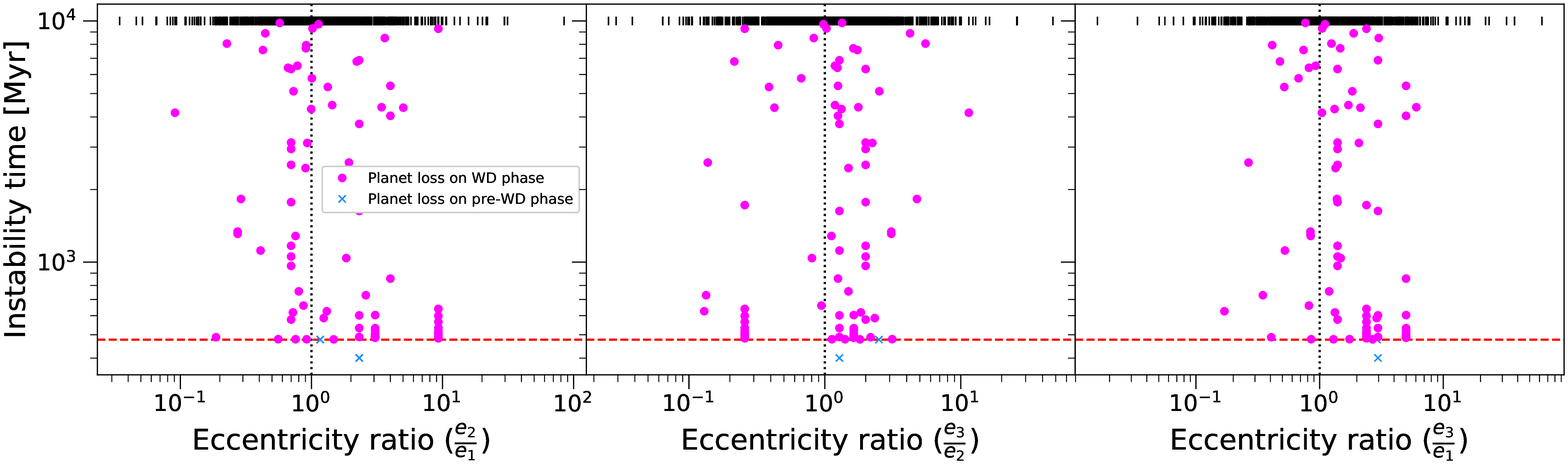} 
\caption{Instability time as a function of the eccentricity ratio for different planet pairs. The black vertical dotted line shows ratio of planets with the same eccentricity value. Symbols are as in Figure \ref{inar}.} 
\label{iner}
\end{center} 
\end{figure*}

We see in Fig. \ref{iner} that the eccentricity ratio of all the planet pairs at which instabilities are happening is rather symmetric with respect to the dotted vertical line that marks the location of the planet pairs with equal eccentricity regarding the covered range of the eccentricity ratios but not with regarding the number of systems. We encounter planet losses in 33 and 43 of the simulations with $\frac{e_2}{e_1}$ $\leq$ 1 and $\frac{e_2}{e_1}$ $>$ 1, respectively, where the distribution of the whole population is 50\,\% with $\frac{e_2}{e_1}$ $>$ 1 and $\frac{e_2}{e_1}$ $\leq$ 1. We have $\frac{e_3}{e_2}$ $>$ 1 in 54 simulations and in 22 for those with $\frac{e_3}{e_2}$ $\leq$ 1 and a similar number of unstable simulations is found for the planet\,1 and 3 eccentricity ratios: 17 with $\frac{e_3}{e_1}$ $\leq$ 1 and 59 with $\frac{e_3}{e_2}$ $>$ 1. Note that the distribution of the eccentricity ratio for those pairs is almost symmetric with a 53\,$\%$ of the systems having ratios $>$ 1 and 47\,$\%$ of them $\leq$ 1. So in the three planet pairs, a major fraction of planet losses take place in systems where the outer planet has a larger eccentricity than the inner planet.

\subsection{Orbital scattering and orbit crossing}

Other dynamical behaviours besides the planet losses analyzed in the previous subsection have the potential to destabilize minor bodies on putative belts that might end up on orbits taking them towards the WD. Such dynamical behaviours are orbit crossing and orbital scattering, and are analyzed in the following. We consider that a system has experienced orbital scattering on the WD phase when at least one of the three planets has a semimajor axis ratio change of at least 5\,$\%$ from the value of the semimajor axis when the WD phase starts. We find that 74 of the 76 simulations where a planet is lost on the WD phase display orbital scattering.  Besides, out of the 76 simulations, 1 simulation have also orbit crossing in the pre-WD phase and 63 simulations have at least one orbit crossing in the WD phase. In addition to the 76 simulations mentioned above, we have 24 simulations that fulfill the scattering criterion in which there is no planet lost. From the 24, 22 simulations do also have orbit crossing on the WD phase. 

If we add all the numbers together we find a total of 100 (8.6\,$\%$) simulations that may contribute to WD pollution either by losing a planet, having orbit crossing and/or orbital scattering on the WD phase.

\begin{figure*}
\begin{center}
\begin{tabular}{c}
\includegraphics[width=18cm, height=14cm]{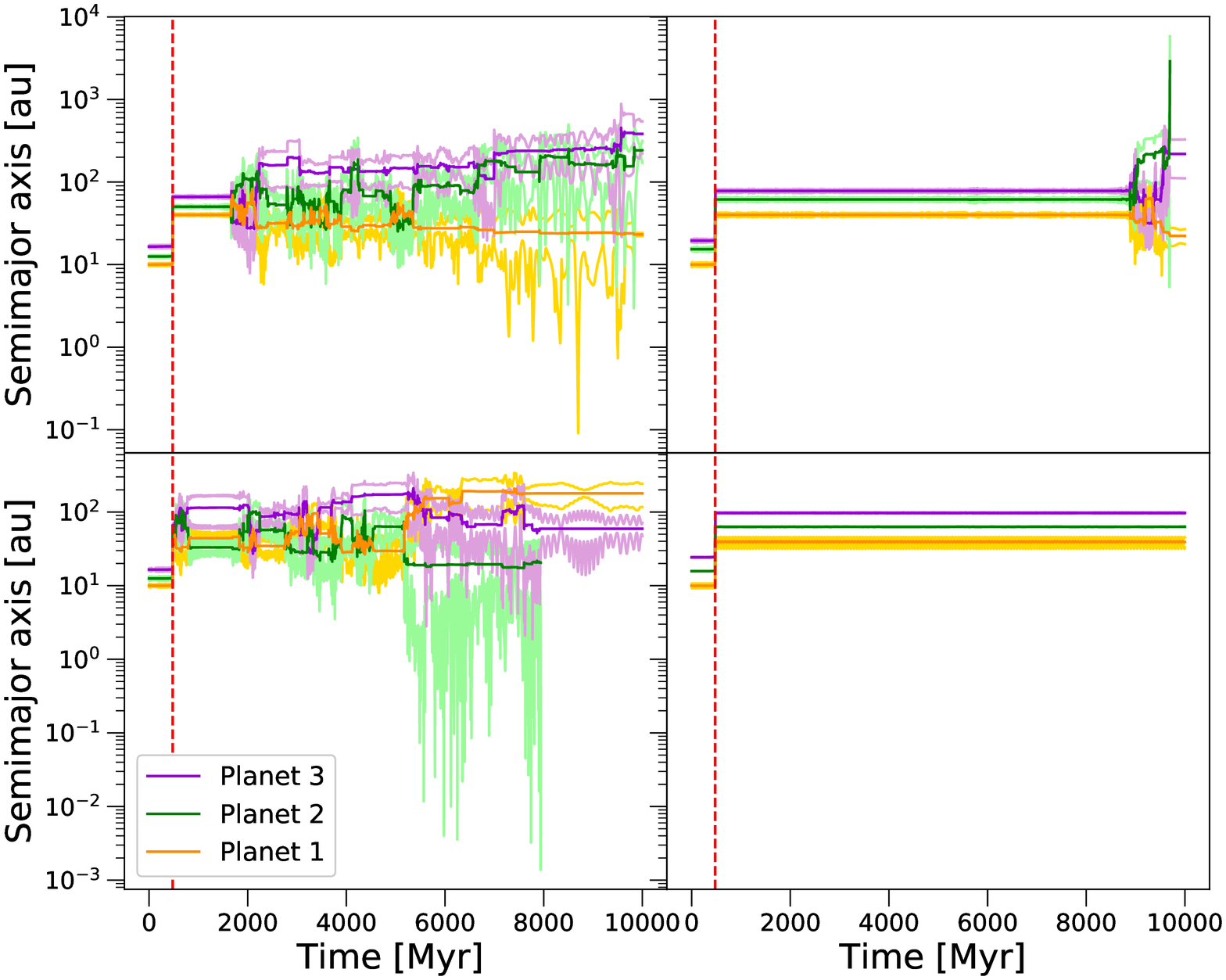} \\
\end{tabular}
\caption{Examples of the semimajor axis evolution of four simulations that show different dynamical behaviours. Solid colours refer to the semimajor of each planet (orange, green and purple for planet\,1, 2 and 3 respectively) and the lighter version of the same colors show the evolution of the pericentre and apocentre of each planet. The red vertical dashed line marks the beginning of the WD phase. In the top left panel we have several orbit crossings and orbital scattering but no planet lost while in the bottom left we have the same but resulting in a planet--star collision around 8 Gyr. The top right panel shows an stable simulation the first 9 Gyr when orbit crossing appear resulting in a planet being ejected. The bottom right panel shows a full stable system during the simulated time.}
\label{exam2}
\end{center} 
\end{figure*}

Illustrative examples of four different dynamical behaviours are displayed in Fig. \ref{exam2}, where we show the semimajor axis evolution along 10 Gyr. The beginning of the WD phase is marked with a red dashed vertical line. In the left panels we have two different simulations of the scaled system  {\it Kepler}--339. In both panels either orbit crossing or orbital scattering are clearly visible in the semimajor axis evolution. The difference between both panels is that in the upper one, no planet is lost despite the chaotic dynamics while in the lower panel,   planet\,2 is lost after several orbit crossings, resulting in a collision with the WD. Note that in both simulations, the pericentre of one planet is reaching a semimajor axis distance $a < $ 1 au due to the eccentricity excitation that the three planets are having caused by the orbit crossing. A more detailed analysis of the planets reaching close distances to the WD is presented in Section \ref{roche}. The top right panel of Fig. \ref{exam2} displays an example of a much less chaotic dynamical behaviour: the scaled system {\it Kepler}--184 where the three planets are dynamically stable until orbit crossing begins after 8.9 Gyr and planet\,2 finally gets ejected at 9.7 Gyr. The lower right panel illustrates the evolution of the scaled system {\it Kepler}--289 that displays a more stable evolution of the three planets, with only a slightly eccentricity excitation in planet\,1 after the formation of the WD, but not enough neither to produce orbital scattering nor orbit crossing along the entire simulated time.   

\subsection{Eccentricity evolution}
\label{eccmax}

\begin{figure}
\begin{center}
\begin{tabular}{cc}
\includegraphics[width=8.5cm, height=6.5cm]{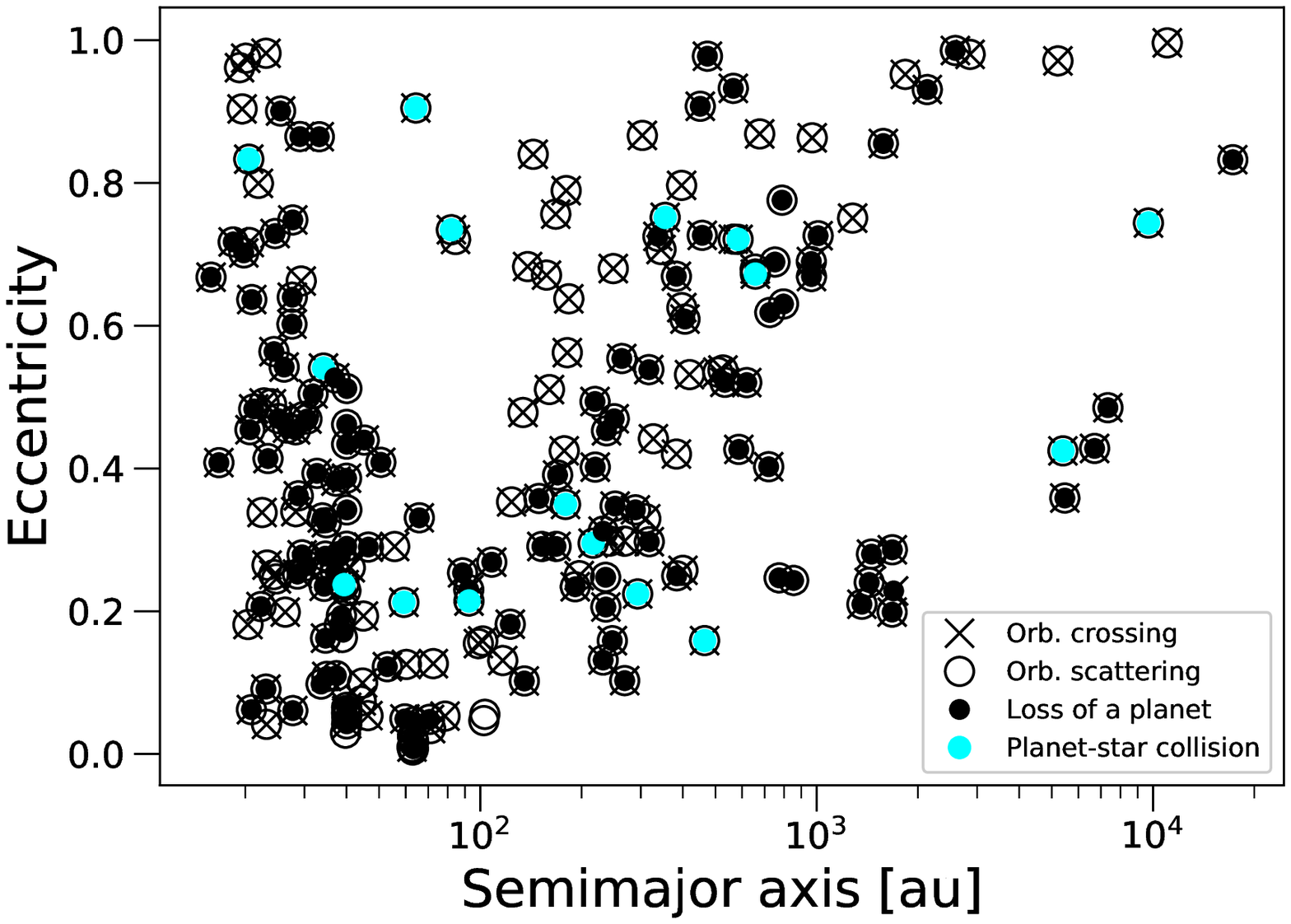} \\
\end{tabular}
\caption{Eccentricities as a function of semimajor axis of the planets that experienced dynamical instabilities: x--symbols are for orbit crossing, black circles for orbital scattering and black dots are for the survivors of the loss of a planet companion and in light blue are the planets surviving the planet--star collision of a companion. Parameters are taken at the end of the simulated time of 10 Gyr.}
\label{finea}
\end{center} 
\end{figure}

As seen in the examples of Fig. \ref{exam2}, the eccentricity of the planets can change as the system evolves. The  excitation of the eccentricity is more effective when there are dynamical instabilities in the simulation, such as planet losses, orbit crossing and/or orbital scattering.  In Fig. \ref{finea} we show the eccentricity as a function of the semimajor axis  for the 148 surviving planets in the 76 simulations that have experienced a planet lost during the WD phase (black dots) and the 72 planets in the 24 simulations that have orbit crossing (x- symbols) and/or orbital scattering (circles) without the lose of any planet. Note that 124 surviving planets also display orbit crossing and 144 have orbital scattering. All the parameters are taken at the end of the simulation (10 Gyr).  Overall the distribution of the planets in the a--e space follows two tails.  Previous works have also shown this behaviour in dynamical simulations \citep[][and Paper\,I]{chatterjee2008,mustill2014}. All the planets that have experienced a dynamical instability cover the whole range of eccentricities from 0 to 0.99. We have highlighted in light blue dots the planets that survive the planet--star collision of their companion. They have final eccentricities $e\geq 0.16$ and up to 0.9.

\begin{figure*}
\begin{center}
\begin{tabular}{c}
\includegraphics[width=18cm, height=6.5cm]{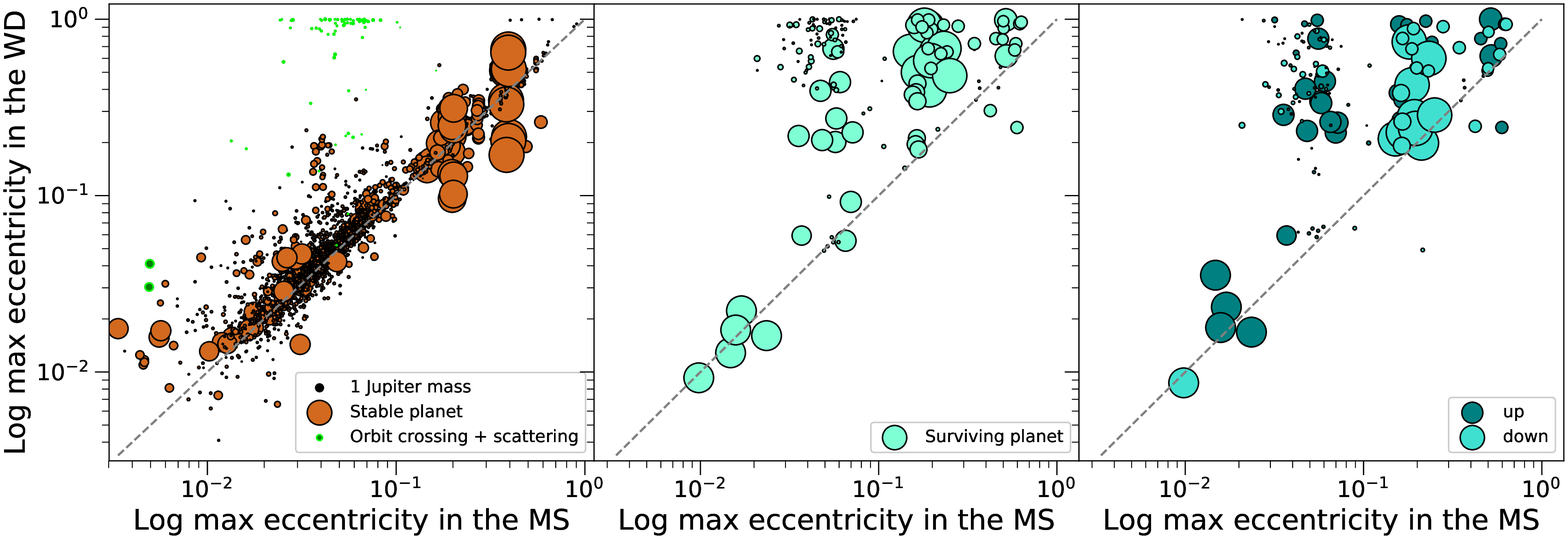} \\
\end{tabular}
\caption{ Right panel: Log-Log scale of the maximum eccentricity at the WD phase reached in the simulations that did not lose any planet as a function of the maximum eccentricity at the MS. The area size of the circles is proportional to the planet mass. Green circles are for planets that experience orbit crossing and orbital scattering at the WD phase. The gray dashed lines represents the 1:1 relation and the size of a Jupiter mass planet is shown in the legend for reference. Middle panel: The maximum eccentricity is obtained until the moment a planet loss in the system. Right panel: The maximum eccentricity is taken after the planet has been lost and until the end of the simulation. The light green mark the surviving planets for which their maximum eccentricity decays after the planet is lost  while the dark green is for planets that increased their maximum eccentricity after the planet loss (referred as down and up in the legend, respectively).}
\label{insmaxe}
\end{center} 
\end{figure*}

\begin{figure}
\begin{center}
\begin{tabular}{c}
\includegraphics[width=8.5cm, height=6.5cm]{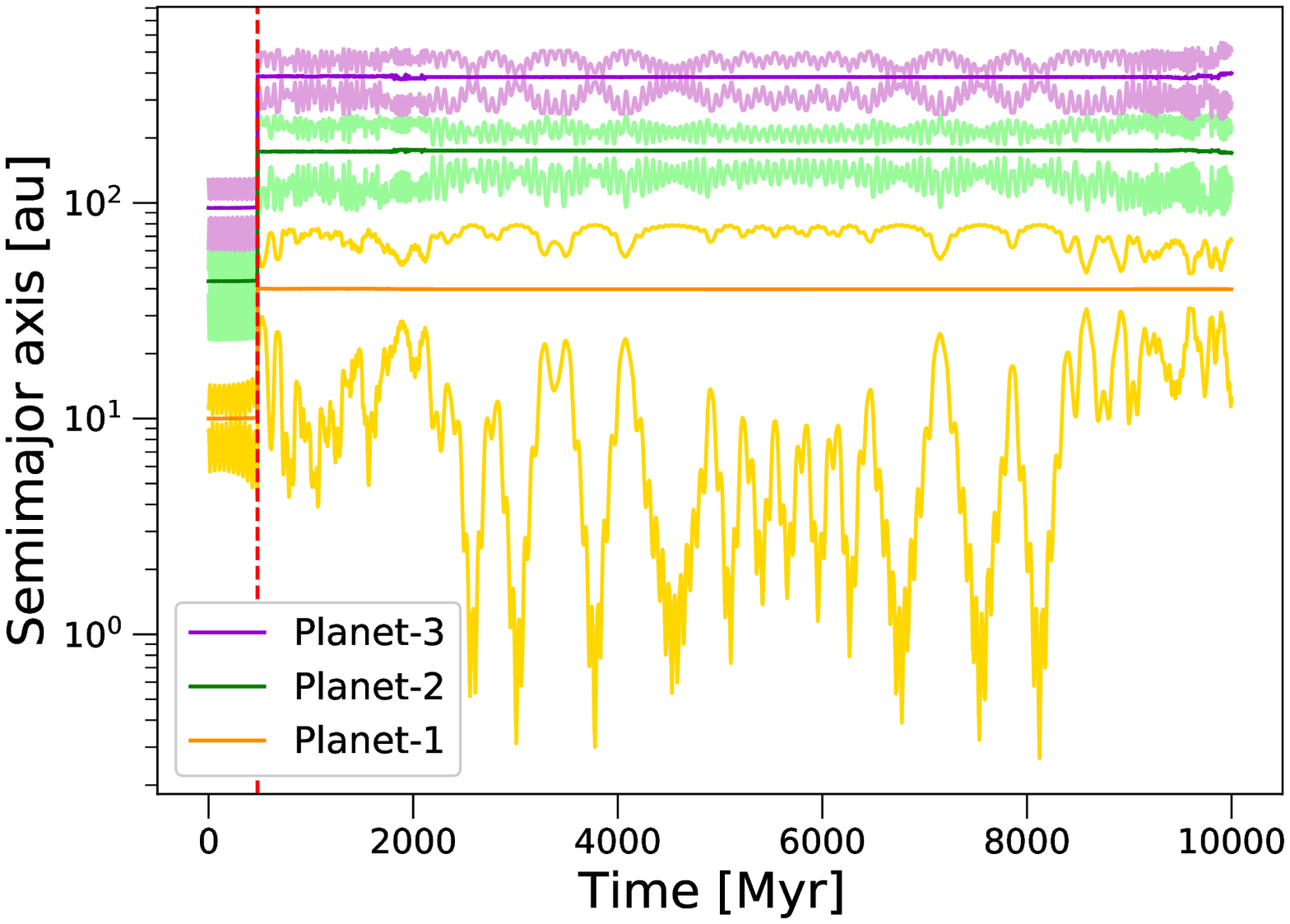} \\
\end{tabular}
\caption{Orbital evolution of the scaled system 61 Vir in one of the simulations.  The pericentre and apocentre of each planet are shown in the lighter version of the colour of the planet. The red vertical dashed line marks the beginning of the WD phase.}
\label{61v}
\end{center} 
\end{figure}

We have done an additional analysis of the eccentricity in the simulations where no planets are lost due to Hill or Lagrange instabilities as the eccentricity might experience changes driven by the stellar mass-loss. In Fig. \ref{insmaxe} we plot the maximum eccentricity on the WD phase as a function of the maximum eccentricity of each planet in the MS. The left panel shows in brown circles each planet with the circle size scaled to the planet mass (for reference we have plotted in the legend the size of a Jupiter mass planet). The green circles show the 72 planets that undergo an orbit crossing and/or orbital scattering in the WD phase without the lose of a planet. Most of the planets follow the dashed line that marks the 1:1 relation meaning that the maximum eccentricity in the MS is similar to the maximum eccentricity on the WD phase for most of them. However, the Hill unstable systems (with orbit crossing) and those that have orbital scattering do separate from the 1:1 relation. We have that 69 out of 72 of the planets in systems dynamically active have a maximum eccentricity in the MS $\leq$ 0.1 (the other 3/72 planets have maximum eccentricity from 0.1 to 0.16 in the MS). The eccentricities of these planets get excited in the WD phase reaching maximum values of $e\geq0.5$ for 56 out of the 72 planets while 16/72 obtain maximum eccentricities e$<$ 0.5. Furthermore, the masses of these eccentric planets are smaller than Jupiter, in the mass range from 0.4 to 41.2 $\mathrm{M}_\oplus$, with only two of them with 366 $\mathrm{M}_\oplus$. Only four planets appear in the upper part of the left panel in Fig. \ref{insmaxe} that are not green circles (with a maximum eccentricity of $\geq 0.9$ in the WD phase). They correspond to the scaled system 61 Vir for which the pericentre of planet\,1 gets very close to the WD in several occasions (see Fig. \ref{61v}).

The eccentricity excitation after a planet is lost is also worth analyzing to check whether the surviving planets continue to be dynamically active. In the middle and right panels of Fig. \ref{insmaxe} and for the 148 surviving planets, we plot the maximum eccentricity on the MS versus the maximum eccentricity at the WD phase taken at two different times. 
The middle panel shows the maximum eccentricity reached between the beginning of the WD phase (478 Myr) up to the moment that their planet companion get lost by Lagrange instabilities. We see that all but 12 planets display a maximum eccentricity in the WD phase larger than that at the MS, an expected result since in 64 out of 76 (84.2\,$\%$) simulations where a planet is lost experience several orbit crossing exciting the eccentricities of the planets before the planet is lost. On the right panel of Fig. \ref{insmaxe}, we show the same as in the middle panel but for the maximum eccentricity reached by the surviving planet from the moment after the planet companion is lost until the end of the simulation (10 Gyr). In this right panel we have plotted the planets whose eccentricities are excited after the planet instability as dark green circles and those for which their maximum eccentricities decays as light green. For the four simulations in which the system is losing more than one planet the maximum eccentricity is taken after the last planet is lost. 88 (out of 148) planets have relaxed after the planet is lost due to Lagrange instability and 60 (out of 148) of them have increased their eccentricity. Regarding the mass of the survival planets, 87/148 have masses $<$ 1 M$_J$ (in the range from 6.7 to 168.3 $\mathrm{M}_\oplus$) and 61/148 have masses $\sim$ 2 -- 20 M$_J$ (between 660.3 and 6255.1 $\mathrm{M}_\oplus$).

Note that we have not included the eccentricity information of the 80 planets that are ejected or suffered collision with the WD, since the nature of the instability sets the maximum eccentricity before the planet loss always $> 0.9$. However, we do have the planet mass information: 49 of them have masses $<$ 124.7 $\mathrm{M}_\oplus$ and 31 have masses $>$ 500 $\mathrm{M}_\oplus$. Regarding the 9 planets that have a planet--star collision, 7 have masses between 6.7 and 124.7 $\mathrm{M}_\oplus$ and two have masses of 2660  $\mathrm{M}_\oplus$.

\subsection{Reaching the Roche radius of the WD}
\label{roche}

Fig. \ref{61v} illustrates one of the cases where a planet gets distances $\leq$ 0.3 au to the WD. Such planets could experience tidal destruction or circularization in their orbits (effects not included in our simulations) and could enter the Roche radius of the WD and collide with it,  hence producing atmospheric pollution. In order to determine which of our systems might be subject to such outcomes we have searched for the simulations with any planet's pericentre entering the Roche radius of the WD, with the Roche radius calculated as
\begin{equation}
a_\mathrm{Roche}=\left(\frac{3\rho_\mathrm{WD}}{\rho_\mathrm{pl}}\right)^{1/3}R_\mathrm{WD}.
\end{equation}
Here, $\rho_\mathrm{WD}$ and $\rho_\mathrm{pl}$ are the densities of the WD and the planet respectively, and $R_\mathrm{WD}$ is the radius of the WD \citep{mustill2014}.

We have obtained 19 (1.6\,$\%$) simulations in which one of the planets' pericentres enters the Roche radius of the WD. Within the 19 simulations, 9 indeed experience a planet--star collision instability (see lower left panel of Fig. \ref{exam2} to see an example of a simulation showing a planet entering the Roche radius before it collides with the WD). The three-planet systems where such collisions happen are the following:  one simulation in the systems EPIC~248545986, {\it Kepler}--226, {\it Kepler}--26, {\it Kepler}--60, {\it Kepler}--445 and two simulations of the system {\it Kepler}--339 and HD~125612. The planet--star collisions happen between 1.7 and 9.3 Gyr. The other 10 simulations where no planet collides with the WD but do enter the Roche radius are:  one simulation in the systems 61~Vir, HD~125612, HD~181433, HD~37124, {\it Kepler}--26; two simulations of {\it Kepler}--60 and three simulations of {\it Kepler}--445. These close approaches happen from 0.02 to 9.1 Gyr after the formation of the WD, where two simulations have planets crossing the Roche radius the first time very early at a time $<$ 100 Myr  and the other 8 have the crossings in times $>$ 1 Gyr. Furthermore, the time difference between the first and the last time a planet enters the Roche radius goes between 0.02 to 8689 Myr. 

Additionally, we have obtained 18 (1.6\,$\%$) more simulations where one of the three planets gets to distances closer than 0.3 au from the WD. In fact, 7 (0.6\,$\%$) of them cross to distances of 10 Roche radii from the WD. From these 18, in 2 simulations the planet that is ejected is having such close distances before the ejection; in 3 simulations a planet survivor is the one that gets close to the WD; and in 13 simulations no planets are lost but the dynamical behaviour of the planets looks like the top left panel of Fig. \ref{exam2}, where several orbit crossings happen throughout the 10 Gyr of simulated time and planet\,1 crosses the 0.3 au threshold at 8.7 Gyr of simulated time for the first time, spanning 16 Myr until the last pericentre passage at 0.3 au. In these simulations, the first time the planet pericentre reaches 0.3 au happens at times larger than 1 Gyr of simulated time (17/18 simulations), even some of them happen at very late times, up to 9.5 Gyr. Only in one simulation the planet reaches the 0.3 au threshold at 50 Myr after the formation of the WD. 

In summary, we obtained a total of 37 (3.2\,$\%$) simulations where at least one planet gets to distances closer than 0.3 au to the WD, with 26 of them (2.2\,$\%$) even reaching the 10 Roche radii threshold, hence having the capability of producing atmospheric pollution, either by direct collision with the WD, tidal disruption of the planet or by destabilization of planetesimals if such belts are located within the path of the planet toward the WD. These simulations correspond to systems where one of the planets reaches eccentricities up to 0.99 on Figure \ref{insmaxe}.

\section{Discussion}
\label{discu}

We find that 2.3\,$\%$ of planets (6.6\,$\%$ of the total simulations) were lost by ejections and planet--star collisions. This number is larger, but comparable, to what we found in Paper\,I (1.5\,$\%$ of simulated planets and 2.9\,$\%$ of the simulations) using the same approach as in this paper but for two-planet systems. On the other hand, \citet{mustill2014} in their study of three-planets (with 1 and 10 Jupiter masses) orbiting a 3 $\mathrm{M}_\odot$ in circular orbits obtained 26\,$\%$ of their planets lost by Hill or Lagrange instabilities, a significantly larger fraction than the one reported here. This discrepancy of results could be explained from the fact that  \citet{mustill2014} included planet losses on the MS phase of the host star but we do not consider them either here nor in Paper\,I  (see \S3). If from \citet{mustill2014} we exclude the simulations with MS instabilities, the number decreases to 13.7\,$\%$ of total planets lost in their simulations, which is still a larger fraction than the one obtained in this work. It is important to mention that nearly 85\,$\%$ of the planet losses in the pre-WD phase in \citet{mustill2014} are happening in systems that have already experienced an instability in the MS and we are excluding these systems from our results.  \citet{mustill2014} found that 14.3\,$\%$ of their simulations started the WD phase as two-planet systems, versus the 0.2\,$\%$ of this paper. After the whole integration (10 Gyr), we ended with 1078/1160 (92.9\,$\%$) simulations as three-planet systems; 78/1160 (6.7\,$\%$) as two-planet systems, 4/1160 (0.3\,$\%$) as one-planet system and we do not have any simulation in which all the three planets are removed. In contrast, the percentage of simulations ending as three, two, one and zero planet systems in \citet{mustill2014} are: 47.2, 28.6, 22.6 and 1.6\,$\%$ respectively. We double the percentage of simulations ending with three planets and find a significantly smaller percentage of simulations with two, one and zero planets.

We find differences in the number of instabilities and the number of survival planets with respect with \citet{mustill2014,mustill2018} besides the fact that we are excluding MS unstable systems and this is related to the fact that we have explored planets with larger separations in terms of $\Delta$. \citet{mustill2014} explored $\Delta$ values in terms of single Hill radius of their planet\,1, from 3 to 18 Hill radii and \citet{mustill2018} have used $\Delta$ in terms of mutual Hill radii from 4 to 12 in all their simulations. We have planets separated from 7.8 to 2859.2 single Hill radii of planet\,1 or 4 to 62.3 in terms of mutual Hill radii of the adjacent pairs. Less than 10\,$\%$ of our simulations are in the range of $\Delta$ explored by \citet{mustill2014,mustill2018}.

In Fig. \ref{indel} we see that the maximum planet separation $\Delta$, in terms of mutual Hill radius, at which instabilities happen on the WD phase, is 27.9 and 25.5 for the planet pairs 1--2 and 2--3 respectively.  \citet{mustill2014} show instabilities up to $\Delta$ $\sim$ 12 single Hill radii ($\Delta\sim$ 7.4 mutual Hill radius). Looking at  Fig. \ref{undel} we observe that our simulations also have a smooth but clear transition between the systems which are unstable during the WD phase and the stable systems, with the planet losses preferably happening in systems with small $\Delta$ values. \citet{mustill2018} also found such a clear transition between the simulations where the first planet is lost on the WD phase and the simulations that remained stable throughout the simulated time for their high-mass planets (100--1000$\mathrm{\,M_\oplus}$), while the 1--30 $\mathrm{M}_\oplus$  planets have a smoother transition between stable and unstable systems. Most of our unstable simulations (51/76; 67.1\,$\%$) have at least one planet with a planet masses $\geq$ 100 $\mathrm{\,M}_\oplus$, and only 25 unstable simulations (32.9\,$\%$) have at least two of the three planets with masses less than 30 $\mathrm{M}_\odot$; hence, our simulations are more comparable to the high-mass simulations of \cite{mustill2018}.

Finally if we look at how the planets are being lost by Hill and Lagrange instabilities, we obtained that 88.8\,$\%$ of the planets lost are ejected, 11.3\,$\%$ collide with the WD and no planet--planet collision happens in this work. In Paper\,I, 85\,$\%$ of the planets were lost by ejections (without counting the ejections produced by the non-adiabatic mass loss), 6.3\,$\%$ had planet--star collisions (which is a similar fraction as the three-planet case) and 8.8\,$\%$ of planets collided each other. \citet{mustill2014} have found also similar percentages: 88\,$\%$ of their planets are lost by ejections, 9\,$\%$ lost by planet--star collisions and only 3\,$\%$ lost by planet-planet collisions on the WD phase. Besides, most of the planets are ejected in the simulations performed by \citet{mustill2018}, especially in the simulations with the planets having masses higher than 100 $\mathrm{M}_\oplus$. This confirms that the planet ejection is the most common type of losing a planet on the WD phase and planet--star, planet--planet collisions are much less common. This outcome is also reproduced by the two-planet cases studied by \citet{veras2013,veras2013b}.

\subsection{Instabilities and cooling age}

\begin{figure}
\begin{center}
\begin{tabular}{c}
\includegraphics[width=8.5cm, height=6.5cm]{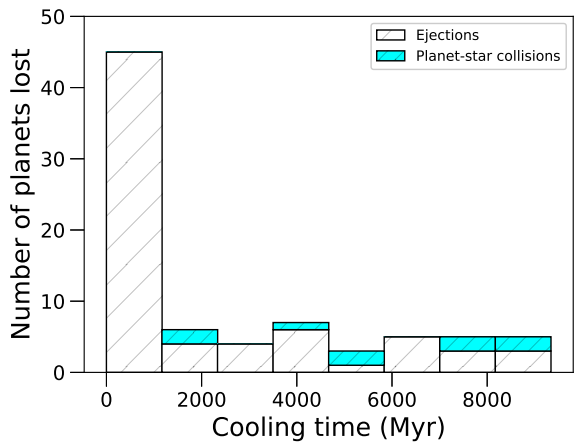} \\
\end{tabular}
\caption{Distribution in the WD cooling time of the dynamical instabilities when a planet is lost on the WD phase due to ejection (white bars) or planet--star collision (light blue bars).}
\label{coolins}
\end{center} 
\end{figure}

A fundamental observable when we try to understand WD pollution is its distribution in time.
In Fig. \ref{coolins}, we have converted the instability time into cooling age of the WD and we show the number of planets lost by ejections (empty bars) and planet--star collisions (light blue bars). Planet ejections occur throughout the whole cooling time with a peak of them happening in the first Gyr of cooling time, after that time, the number of planet ejections happen at a constant rate up to 10 Gyr. Furthermore, all but two planet--star collisions are within 4.6 to 8.8 Gyr. In detail,  we have 24/3480 (0.7\,$\%$) planet losses in the first 100 Myr; 20 (0.6\,$\%$) happening between 100 Myr and 1 Gyr of WD cooling time and 36 (1.0\,$\%$) happening after 1 Gyr.   

We find our three-planet systems become unstable at very late ages of cooling time, especially the planet--star collisions, which may explain the atmospheric pollution found in very late WDs \citep{farihi2009,dufour2017b,hollands2018}. The mean and median of cooling time when the planet losses happen in this work is 10$^{9.4}$ and 10$^{9.1}$ yr respectively. In contrast, \citet{mustill2014} found that the median age of the instabilities producing the first planet losing at WD phase is 10$^{7.9}$ yr for planets with one Jupiter mass, while 10$^{6.5}$ yrs for planets with 10 Jupiter mass. We do find as well as \citet{mustill2014} that most of the ejections happen earlier than the planet--star collisions \citep[see e.g. Fig. 7 and 10 of][]{mustill2014}.

In order to understand the role of the planet mass on the final outcome and the possible influence on the WD pollution, we have divided the 98 simulations with orbit crossing and/or loss of a planet in two groups according to the planet mass. The first group includes the simulations where at least one of the three planets have mass $\geq$ 100 $\mathrm{M}_\oplus$ (high mass planets simulations). The second group  have planets with masses between 1--100 $\mathrm{M}_\oplus$ (low mass planets simulations). Both onsets have 47/98 and 51/98 (48 and 52\,$\%$) simulations, respectively. Interestingly, 45/47 simulations of the onset of low mass planets are having two or the three planets in the mass range of 1--30 $\mathrm{M}_\oplus$. 

The low mass planets simulations have a mean and median cooling time of planet losses happening at 5.1 and 4.9 Gyr respectively, while those with high mass planets peak at earlier times with a mean of 958 Myr and a median of 123 Myr. The same behaviour is found regarding simulations with orbit crossings (with and without planet losses), low mass planets have a mean and median of 2.7 and 1.3 Gyr respectively that compare to the 704 and 25 Myr mean and median respectively of the high mass planets. We confirm previous results (\citet{mustill2014} and Paper\,I) that planetary architectures involving Jovian planets are getting destabilized earlier than those systems having super Earth--Neptune mass planets. 

\begin{figure*}
\begin{center}
\begin{tabular}{c}
\includegraphics[width=16cm, height=7cm]{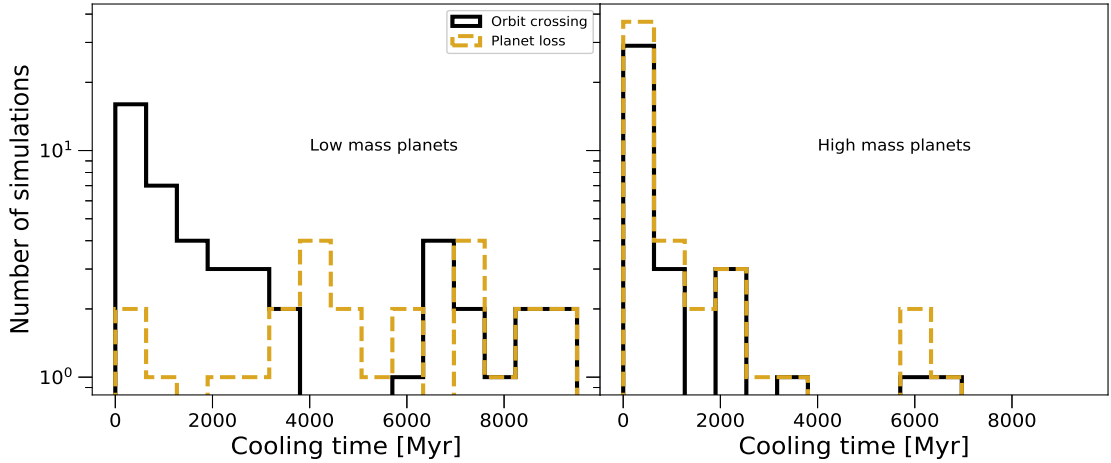} \\
\end{tabular}
\caption{Distribution of simulations having a dynamical instability (orbit crossing as a black solid line and planet loss as a gold dashed line) along the WD cooling time. In the left panel we show simulations with planet masses in the range of 1--100 $\mathrm{M}_\oplus$ (low mass case), while the right panel depicts simulations with at least one planet having masses $\geq$ 100 $\mathrm{M}_\oplus$ (high mass case). If the simulation has more than one orbit crossing or planet losses, then we record the time of the first orbit crossing and/or the first planet loss.}
\label{coolcross}
\end{center} 
\end{figure*}

In Fig. \ref{coolcross} we have built  WD cooling time histograms of the simulations including as a solid black line orbit crossings and in gold dashed line planet losses by Lagrange instabilities. We take into account the time of the first orbit crossing when several of them happen in the same simulation, and, we record the time of the first planet loss (in the case of several). The low mass planet sample is in the left panel and simulations with at least one high mass planet are plotted the right panel. 

Both onsets (low and high mass planets) display a peak of orbit crossings happening in the first Gyr of the WD cooling time, with time the number of orbit crossing decreases for the high mass case (right) while after a gap remains a low rate for the low mass sample. The planet losses remain constant for low mass planets along the entire cooling time but peak for the high mass regime in the first Gyr decreasing rapidly and showing a second little burst at 6 Gyr. A very similar dynamical behaviour in cooling time is  found in \citet[see e.g. lower panels in Fig. 4]{mustill2018}. Our simulations harbouring planets with masses larger than 100 $\mathrm{M}_\oplus$ are dynamically active the first Gyr of cooling time: the number of instability events rapidly decreases with no events happening beyond 6.4 Gyr. In the simulations with planet masses $\leq$ 100 $\mathrm{M}_\oplus$ (most of them with masses $<$ 30 $\mathrm{M}_\oplus$) the dynamical instabilities happen along the entire simulated cooling time and events occur up to 9.5 Gyr.

\citet{bonsor2011,frewen2014} studied the interaction of one-planet systems with a planetesimal belt, concluding that eccentric (e$>$ 0.4) and low mass planets can contribute to send material efficiently toward the WD. We have seen in Section \ref{eccmax} that our Hill and Lagrange unstable simulations, as well as the simulations with orbital scattering have planets that achieved large eccentricities due to the dynamical interaction among them. Within the 24 simulations (72 planets) where no planets are lost but do have orbit crossing and scattering,  we have that 56/72 (77.8\,$\%$) of the planets, 1.6\,$\%$ with respect to the total planets simulated, are having eccentricities $\geq$ 0.4 and planet masses between 0.4 to 30.5 $\mathrm{M}_\oplus$. Indeed, these planets are in the mass range that \citet{mustill2018} found as the most efficient deliverers of planetesimals to the WD for Gyr time scales of cooling time. 

Additionally, by considering the 148 surviving planets and their maximum eccentricities after the planetary systems become two- or one-planet systems from the time of the planet loss until the end of the simulation we found that 82/148 (55.4\,$\%$, or 2.4\,$\%$ with respect to the total planets simulated) are having maximum eccentricities larger than 0.4 and 37 of them have low masses. These planets could be contributing to send material towards the WD while Jovian mass planets preferentially eject the material from the system.

In summary, from the 300 planets (100 simulations, 8.6\,$\%$ of total simulations) involved in dynamical instabilities at the WD phase, such as losing planets and/or having orbit crossing, scattering  we have that 186 planets (5.3\,$\%$ with respect to the total planets simulated) would be efficient deliverers of planetesimals to the WD, based on their masses. The fraction of planets decreases to 140 (4.0\,$\%$) if we add the eccentricity restriction $e>0.4$.

\subsection{Planet approaches to the WD}

In this work we find that 1.2\,$\%$ of the planets simulated reach pericentre distances  $\leq$ 0.3 au from the WD. We obtain similar results to previous works (0.1\,$\%$ in Paper I and 0.2\,$\%$ \citet[][see e.g. Fig. 8]{mustill2014}). The slight increase of planets reaching distances $\leq$ 0.3 au to the WD between the two-planet case and this work is due to the fact that we have obtained more simulations with orbit crossing and/or scattering (7.4\,$\%$) than Paper\,I (1.3\,$\%$). Fig. \ref{insmaxe} (see Section \ref{eccmax}) confirm the latter statement by showing that the maximum eccentricity reached in the WD phase (up to $e>0.9$) is much higher than the maximum value in the MS in simulations where planet losses, orbital scattering and/or orbit crossing is present. Thus, more orbit crossing and scattering in the three-planet case increase the likelihood of having planets with high eccentricity and in consequence their planet pericentres at short distances from the WD.

If we look at how many of the planets cross the Roche radius of the WD we obtain a 0.5\,$\%$ of the planets simulated that have masses in the range from 6.7 to 124.7 $\mathrm{M}_\oplus$ (only six planets have masses $>$ 800 $\mathrm{M}_\oplus$). Furthermore, we have 0.2\,$\%$ more planets with masses $\leq$ 30 $\mathrm{M}_\oplus$ and one planet with 750 $\mathrm{M}_\oplus$ that cross the pericentre distance of 10 Roche radii where the planetary bodies can still be influenced by tidal forces \citep{veras2019b}. These planets are very interesting since our simulations provide a mechanism to send Neptune-, Saturn- and Jupiter-like planets to very close distances to the WD where they can be exposed to the evaporation of their atmospheres, as the case of the inferred planet orbiting WD J091405.30+191412.25 \citep{gansicke2019}, or disrupted to their cores as proposed by \citet{manser2019} to explain the origin of the orbiting body in the gas disk around SDSS J122859.93+104032.9. Also, the tidal forces in such close distances of the WD can produce the self-disruption of ice giants or circularization of gas giants \citep{veras2019}, effects not included in our simulations but deferred to a future work. We confirm nonetheless that planets reaching close distances from the WD are rare in two- and three-planet systems.

\section{Conclusions}
\label{conclu}

We have evolved 135 three-planet systems (with 10 simulations per system configuration) during 10 Gyr. The planets are orbiting a 3 $\mathrm{M}_\odot$ host star, from the MS along the RGB and AGB phases and finally ending as a WD. In order to constrain the infinite physical and orbital parameter space of the planetary systems we selected observed systems and explored a wider parameter space than previous works.  By removing 190 simulations which have dynamical instabilities in the MS phase (planet losses and orbit crossing) due to having planet pairs in mean motion commensurabilities, we finally considered 1160 simulations (3480 planets simulated) for our statistical analysis. 

We found that 76 simulations (6.6\,$\%$) lose 80 planets (2.3\,$\%$) by Lagrange instabilities on the WD phase, mainly by ejections with only 9 planets colliding with the WD. The number of simulations/planets increases by adding other dynamical instabilities, such as orbit crossing and orbital scattering with no planet losses, making a total of 100 simulations (8.6\,$\%$) dynamically active in the WD phase. In Table \ref{tabins} we list the observed planetary systems for which their scaled versions have presented dynamical instabilities in our simulations. The first column gives the name of the observed system, the second the percentage of the simulations of such system that are unstable on the WD phase and the following columns give the type of instability present in at least one of the simulations.

Although we can not explain the estimated percentage of metal polluted WD (25--50\,$\%$) with dynamical instabilities using these three-planet systems, we do have a slight increase of simulations which can produce pollution with respect to the 3.2\,$\%$ found in the two-planet systems (Paper\,I). Then, it is possible than multiple-planet systems with four and more planets would continue to increase the rate of simulations contributing to the WD pollution. Simulations with four and more planets will be performed in a future study.

Additionally, our three-planet simulations show that dynamical instabilities are prone to happen in the first Gyr of the WD cooling time and then decreases for late WD ages. This outcome is in correspondence with \citet{hollands2018} who found the slow decay of the material reservoir that pollutes WD with an e-folding timescale of 1 Gyr. Besides, we find that low mass planets can experience instabilities starting after 1 Gyr of cooling time, while Jovian planets typically have earlier instabilities, in agreement with \citet{mustill2014,mustill2018}. 

Our three-planet simulations also show a mechanism to send planets toward the host star and some of these planets can even cross the Roche radius of the WD (1.6\,$\%$ of our simulations within the 2.2\,$\%$ with pericentre passages $\leq$ 10 Roche radii), where a direct collision to the WD surface, photoevaporation of the planet atmosphere, disruption or circularization can happen, hence producing a dust/gas disk that eventually may pollute the WD \citep{manser2019,gansicke2019,veras2019,veras2019b}. Furthermore, our $\sim$ 1.6\,$\%$ of simulations where planets have pericenter passages within the Roche radius is in agreement to the 1--3\,$\%$ of polluted WD showing the evidence of a circumstellar debris disk \citep{wilson2019} and to the 0.04--1.11\,$\%$ showing a gaseous component \citep{manser2020}.

\begin{table}
\begin{center}
\caption{Three-planet systems with dynamical instabilities in the WD phase. Y-symbol marks the dynamical instability of each column happening in at least one simulation of the scaled system. Note that the column referring to planet-planet collision does not appear in the table since no systems have such instability.}
\label{tabins}
\resizebox{.47\textwidth}{!}{%
\begin{tabular}{l c c c c c }
\noalign{\smallskip} \hline \noalign{\smallskip}
\multicolumn{2}{c}{} & \multicolumn{4}{c}{Dynamical instabilities during the WD phase} \\
System & $\%$ & Ejection & Planet-star  & Orbit & Orbital  \\
 & unstable & & collision  & crossing & scattering\\
\noalign{\smallskip} \hline \noalign{\smallskip}
 47~Uma & 100 & Y & -  & Y & Y \\
 EPIC~248545986 & 100 & - & Y & Y & Y  \\
 HD~125612 & 100 & Y & Y & Y & Y  \\
 HD~181433 & 90 & Y & - &  Y  & Y  \\
 HD~37124 & 85 &  Y & - &  Y & Y   \\
 {\it Kepler}-138 & 20 & - & - &  Y & Y    \\
 {\it Kepler}-184 & 20 & Y & - &  Y & Y    \\
 {\it Kepler}-191 & 10 & - & - &  Y & Y   \\
 {\it Kepler}-217 & 30 & Y & - &  Y & Y    \\
 {\it Kepler}-226 & 78 & Y & Y &  Y & Y   \\
 {\it Kepler}-26 & 60 & Y & Y &  Y & Y   \\
 {\it Kepler}-271 & 10 & - & - &  Y & Y  \\
 {\it Kepler}-279 & 30 & Y & - &  Y & Y   \\
 {\it Kepler}-289 & 20 & - & - &  - & Y  \\
 {\it Kepler}-30 & 10 & Y & - &  - & Y  \\
 {\it Kepler}-31 & 100 & Y & - &  - & Y   \\
 {\it Kepler}-339 & 100 & Y & Y &  Y & Y    \\
 {\it Kepler}-350 & 30 & Y & - &  Y & Y   \\
 {\it Kepler}-359 & 20 & Y & - &  Y & Y   \\
 {\it Kepler}-445 & 70 & Y & Y &  Y & Y   \\
 {\it Kepler}-60 & 50 & Y & Y &  Y & Y   \\
 {\it Kepler}-65 & 50 & Y & - &  Y & Y   \\
\hline
\end{tabular}}
\end{center}
\end{table}

\section*{Acknowledgements}
We thank the referee for carefully reading the manuscript and giving helpful comments to improve this work. This research has made use of the NASA Exoplanet Archive, which is operated by the California Institute of Technology, under contract with the National Aeronautics and Space Administration under the Exoplanet Exploration Program.  This research has made use of the SIMBAD database, operated at CDS, Strasbourg, France.  E.V. and R.M. acknowledge support from the `On the rocks II project' funded by the Spanish Ministerio de Ciencia, Innovaci\'on y Universidades under grant PGC2018-101950-B-I00. MC, RM and EB thank CONACyT for financial support through grant CB-2015-256961. A.J.M. acknowledges support from the project grant 2014.0017 `IMPACT' from the Knut \& Alice Wallenberg Foundation, and from the starting grant 2017-04945 `A unified picture of white dwarf planetary systems' from the Swedish Research Council. We are grateful to Rafael Gerardo Weisz and Francisco Prada for their help with the automation of the processes and the use of the cluster.

\section*{Data availability}

The data underlying this article will be shared on reasonable request to the corresponding author.




\bibliographystyle{mnras}
\bibliography{bib} 








\bsp	

\label{lastpage}
\end{document}